\documentclass[a4paper,11pt]{article}
\pdfoutput=1
\usepackage{jcappub}
\usepackage{amsthm,graphicx,multirow}
\usepackage{epsfig}
\usepackage{latexsym, amssymb} 
\usepackage{amsmath}
\usepackage{epstopdf}
\def\beq{\begin{equation}}
\def\eeq{\end{equation}}
\def\br{\begin{eqnarray}}
\def\er{\end{eqnarray}}
\def\benu{\begin{enumerate}}
\def\efnu{\end{enumerate}}

\def\l{\left}
\def\r{\right}

\begin{document}
\title{Probing features in inflaton potential and reionization history with future CMB space observations}
\author[1]{Dhiraj Kumar Hazra,}
\author[2,3]{Daniela Paoletti,}
\author[4,2,5,3]{Mario Ballardini,}
\author[2,3]{Fabio Finelli,}
\author[6,7]{Arman Shafieloo,}
\author[1,8,9,10]{George F. Smoot,}
\author[11,12]{Alexei A. Starobinsky}
\affiliation[1]{AstroParticule et Cosmologie (APC)/Paris Centre for Cosmological Physics, Universit\'e
Paris Diderot, CNRS, CEA, Observatoire de Paris, Sorbonne Paris Cit\'e University, 10, rue Alice Domon et Leonie Duquet, 75205 Paris Cedex 13, France}
\affiliation[2]{INAF/IASF Bologna, via Gobetti 101, I-40129 Bologna, Italy}
\affiliation[3]{INFN, Sezione di Bologna, via Irnerio 46, I-40127 Bologna, Italy}
\affiliation[4]{Department of Physics \& Astronomy, University of the Western Cape, Cape Town 7535, South Africa}
\affiliation[5]{Dipartimento di Fisica e Astronomia, Universit\`a di Bologna, Viale Berti Pichat, 6/2, I-40127 Bologna, Italy}
\affiliation[6]{Korea Astronomy and Space Science Institute, Daejeon 34055, Korea}
\affiliation[7]{University of Science and Technology, Daejeon 34113, Korea}
\affiliation[8]{Institute for Advanced Study \& Physics Department, Hong Kong University of Science and Technology, Clear Water Bay, Kowloon, Hong Kong}
\affiliation[9]{Physics Department and Lawrence Berkeley National Laboratory, University of California, Berkeley, CA 94720, USA}
\affiliation[10]{Energetic Cosmos Laboratory, Nazarbayev University, Astana, Kazakhstan}
\affiliation[11]{Landau Institute for Theoretical Physics RAS, Moscow, 119334, Russian Federation}
\affiliation[12]{Kazan Federal University, Kazan 420008, Republic of Tatarstan, Russian Federation}
\emailAdd{dhiraj.kumar.hazra@apc.univ-paris7.fr, paoletti@iasfbo.inaf.it, mario.ballardini@gmail.com, 
finelli@iasfbo.inaf.it, shafieloo@kasi.re.kr, gfsmoot@lbl.gov, alstar@landau.ac.ru} 

\abstract 
{We consider the prospects of probing features in the primordial power spectrum with future 
Cosmic Microwave Background (CMB) polarization measurements.
In the scope of the inflationary scenario, such features in the spectrum
can be produced by local non-smooth pieces in an inflaton potential
(smooth and quasi-flat in general) which in turn may originate from fast
phase transitions during inflation in other quantum fields interacting
with the inflaton. They can fit some outliers in the CMB temperature power spectrum 
which are unaddressed within the standard inflationary ${\mathrm{\Lambda}}$CDM model. 
We consider Wiggly Whipped Inflation (WWI) as a theoretical framework leading to improvements in
the fit to the Planck 2015 temperature and polarization
data in comparison with the standard inflationary models, although not at a 
statistically significant level. 
We show that some type of features in the potential within the WWI models, leading to oscillations
in the primordial power spectrum that extend to intermediate and small scales can be constrained with high confidence
(at 3$\sigma$ or higher confidence level) by an instrument as the Cosmic ORigins Explorer (CORE).
In order to investigate the possible confusion between inflationary features and footprints from the
reionization era, we consider an extended reionization history with monotonic increase of free electrons with decrease in redshift. We discuss the present
constraints on this model of extended reionization and future predictions with CORE. We also project, to what extent, this extended reionization can create confusion in identifying 
inflationary features in the data.}

\maketitle

\section{Introduction}

With its nine frequency channels, Planck has probed the CMB anisotropy pattern with unprecedented accuracy
and has provided the tightest constraints on cosmological parameters~\cite{Planck:2015Param,Planck:2015Like} and on the 
primordial power spectrum (PPS) from the simplest models of inflation~\cite{Planck:2015Inf}. 
The Planck 2015 data are statistically consistent with a power law form for the PPS~\cite{Planck:2015Inf} 
and with Gaussian perturbations~\cite{Planck:2015NG}, which are both compatible with the predictions 
of standard single field slow roll inflation models.

Although not at a statistically significant level, an improved fit to the temperature power spectrum by 
Planck~\cite{Planck:2013Inf,Hazra:2013BroadRecon,Hazra:2014PPSPlanck,Planck:2015Inf} 
can be obtained by features in the PPS as provided by temporary violation of the slow-roll 
condition during inflation produced by non-smooth pieces in an inflaton potential (smooth and
quasi-flat in general). The Wiggly Whipped Inflation (WWI) has provided a framework for classes of model with 
primordial features which improves the fit to the Planck 2015 data with respect to the power law model~\cite{Hazra:2014WWI,HSSS:2016}. 
These features can be localized on certain scales or they can represent wiggles over a wide range of cosmological scales.
Within the WWI framework, we consider inflationary phase transitions where 
the scalar field rolls from a steeper potential into a nearly flat potential through a discontinuity in the potential or in its derivative. 
We obtained five types of features that provide a $\Delta\chi^2\sim12-14$ improvement, compared to the power law, in the fit to the Planck data, 
with the addition of 2-4 extra parameters depending on the model. These features help improve the fit to the Planck 2015 temperature 
and polarization anisotropies separately and in combination. Four types of features come from the model where we have a (smoothed) discontinuity in the potential.
The remaining type is instead generated by a discontinuity in the first 
derivative of the potential. The baseline potential used in the latter case is the $\alpha$-attractor model~\cite{alpha_attractor}, 
in a limit where it reduces to Starobinsky $R + R^2$ model~\cite{Starobinsky:1980te}.
Since a large fraction of inflation models belong to a universal class where spectral tilt and tensor 
amplitude follow the same function of {\it e-folds}, it is interesting to use such potential 
in a framework for generating features. Since the release of WMAP data, different features have been tested 
with the data, such as features generated with a step in the inflaton potential or in its
first derivative~\cite{S92,step-models,Hazra:2010Step}, oscillations in the inflaton potential~\cite{oip} or through an inflection point~\cite{Jain:2009PI}~\footnote{
Another possibility to obtain localized oscillatory features in the
primordial scalar power spectrum which we do not consider in this paper is
a transient change in the inflaton effective sound speed~\cite{Achucarro:2013cva} that may
occur during multiple inflation.}.
Several reconstruction of the PPS from the CMB have demonstrated~\cite{reconstruction-all}
the hints of suppression of power and oscillations at large and intermediate scales ($\ell\lesssim 40$) and sharper oscillations that continues to small scales. 
As mentioned before, given the uncertainties in the CMB data associated at different scales, while these features are not statistically significant,
they are interesting because of following three reasons: firstly, some of these features are consistently present since WMAP 
observations~\cite{Peiris:2003ff,Komatsu:2008hk,Komatsu:2010fb,Hinshaw:2012aka}; secondly, Planck temperature and polarization observations 
jointly support these features; and finally, within a single framework of inflation, WWI, these three classes of features can be generated
at the required cosmological scales~\cite{HSSS:2016}.

The main goal of this paper is to investigate what a future concept for space mission dedicated to CMB polarization 
can tell about the existence of features as generated by a temporary violation of the slow-roll condition during inflation.
We focus on space missions because only space experiments can observe the full 
sky and therefore access the largest angular scales.  
We consider for our forecast the CORE proposal~\cite{coresite,core:inst,Delabrouille:2017rct,core:cosmoparam,core:inf} as a 
concept for a future CMB polarization dedicated space mission~\cite{Matsumura:2013aja,Matsumura:2016sri,Kogut:2011xw}. The study presented here is complementary to the CORE 
forecasts for primordial features presented in~\cite{core:inf}. 

Along with forecasting the capabilities of a future mission as CORE regarding primordial features, 
we will also investigate the current constraints on the 
reionization history and the corresponding CORE forecasts. 
Beyond the interest in the process of reionization per se, there is an additional reason to investigate this specific topic 
in this paper. 
Indeed, large scale features in the PPS are degenerate to some extent with extended reionization histories in the CMB polarization spectrum: 
it is therefore important to investigate the effect of 
this degeneracy in the constraints on primordial features, in particular in the perspective of future high accuracy polarization data. 
There are still grey areas in the knowledge of the process of reionization: in particular, the details of the transition from 
neutral hydrogen to a fully reionized medium (for $z>6$) are still unknown, and nevertheless they affect the large 
angular scale CMB polarization pattern and have consequences on the determination of cosmological parameters.

Attention has been already paid in the past~\cite{mortonson} to the degeneracy between primordial features and reionization . 
Using the Principal Components as basis for parametrizing general reionization histories, it has been
demonstrated that complex reionization histories~\cite{PCA} can reduce the 
significance of primordial features for a non-standard inflationary model with a step in a quadratic potential~\cite{mortonson}. 
Keeping this in mind, in this paper we use a one parameter controlled reionization 
history that allows more freedom than a Tanh step (that is usually assumed 
in the Einstein-Boltzmann codes), but at the same time does not allow a complete free form reionization history as in PCA.
Also this parametrization does not allow unphysical reionization histories.
We consider a monotonic increase in neutral hydrogen fraction with redshift and 
we provide the constraints with Planck 2015 data and the forecast for CORE capabilities. 
We also discuss the degeneracy between the reionization history and a particular shape 
of large scale power spectra supported by Planck data.

The paper is organized as follows: in section~\ref{sec:inf} we review the WWI potential. We first summarize the features that are supported by Planck 2015 data~\cite{HSSS:2016}, 
we introduce a minor change in the WWI parametrization with respect to previous treatments.
In section~\ref{sec:wwiforecast} we provide all the necessary details for the analysis and present the results. 
In the reionization dedicated part~\ref{sec:reion}, we discuss the different parametrizations, both the commonly used model and the one utilized in this paper.
We present the methodology and priors, together with the constraints we obtained. 
In section~\ref{sec:degen} we provide our study of the degeneracies between the primordial power spectrum and extended reionization histories. 
We conclude in section~\ref{sec:conclusions}. 

\section{Inflation and the primordial power spectrum}~\label{sec:inf}

We start with the discussion of the WWI framework, in a slightly modified form with respect to previous treatments, and the Planck 2015 results.
\subsection{Wiggly Whipped Inflation}

In the WWI framework the potential is described by: 
\begin{equation}
V({\phi})=V_{S}(\phi)+ V_{R}(\phi),~\label{eq:equation-WWI-basic}
\end{equation}
where, $V_{R}$ is the steep part which merges into the slow roll part $V_{S}$ with or without a discontinuity.
This steep to flat phase transition in inflation allows features in the PPS on cosmological scales.
In the literature~\cite{S92,Linde:1998OpenInf,Linde:1999OpenInf2,CPKL03,JSS08,JSSS09,Bousso:2013uia} the possibility of a phase transition in the inflaton potential has been discussed within different frameworks. The WWI one, in particular, allows the potential and/or its derivatives to have discontinuities.

\subsubsection{Discontinuity in the potential}
We use the WWI potential as provided below in Eq.~\ref{eq:equation-WWI}: 
\begin{equation}
V({\phi})=V_{0} \l[\l(1-\l(\frac{\phi}{\mu}\r)^{p}\r)+\Theta(\phi_{\rm T}-\phi)\l(\gamma (\phi_{\rm T}-\phi)^{q}+\phi_{0}^q\r)\r],~\label{eq:equation-WWI}
\end{equation}
where we note that $V_{S}(\phi)=V_{0} \l(1-\l(\frac{\phi}{\mu}\r)^{p}\r)$ has 2 parameters, $V_{0}$ and $\mu$. $\mu$ and the index $p$ determine the spectral tilt $n_{\rm s}$
and the tensor-to-scalar ratio $r$. 
We choose the values $p=4$ and $\mu=15~{\rm M_{PL}}$ such that $n_{\rm s}\sim0.96$ and $r\sim{\cal O}(10^{-2})$ (as in~\cite{Efstathiou:2006ak,Hazra:2010Step}). 
With respect to the previous treatment, here we change a bit the steep part in the potential $V_{R}(\phi)$. It is composed by two independent part: $\gamma (\phi_{\rm T}-\phi)^{q}$ 
which generates the whipped suppression~\cite{Hazra:2014WI} through the fast roll, and $\phi_{0}^q$ which introduces the wiggles through the discontinuity. 
Hence, the potential discussed in~\cite{HSSS:2016} is nested within this potential. The extent of the potential discontinuity is $\Delta V = V_0 (\phi_0)^q$.
We use $q=2$ as in~\cite{HSSS:2016}.
The transition and discontinuity happen at the same field value $\phi_{\rm T}$ just like the old potential.
In this case to have a featureless PPS it is necessary to have satisfied both $\gamma=0$ and $\phi_{0}=0$.
Separately, $\phi_{0}=0$ reduces the potential to Whipped Inflation form and $\gamma=0$ reduces the 
potential to a form where we do not have large scale suppression in the PPS but we generate localized 
and non-local wiggles. The Heaviside Theta function $\Theta(\phi_{\rm T}-\phi)$ is modeled numerically as usual by a Tanh step
($\frac{1}{2}\l[1+{\tanh}[{(\phi_{\rm T}-\phi)}/{\delta}]\r]$) and thereby introduces a new extra parameter $\delta$. 
Note that $\phi$, $\phi_{\rm T}$ and $\phi_0$ are all measured in units of the
reduced Planck mass.

With `WWI potential', we shall refer to the potential in Eq.~\ref{eq:equation-WWI}. 
The main advantage of using WWI potential is that within a single potential we 
can generate a multitude of features that are supported by the CMB data. 
It acts as a generic model to avoid the use of different potentials for parameter 
estimation. 

\subsubsection{Discontinuity in the derivative of the potential}
With `WWI$'$ potential' we shall refer to the potential with a discontinuity in the derivative which reads as follows:
\begin{equation}
V({\phi})=\Theta(\phi_{\rm T}-\phi) V_{0} 
\l(1-\exp\l[-\alpha\kappa\phi\r]\r)+\Theta(\phi-\phi_{\rm T}) 
V_{1}\l(1-\exp\l[-\alpha\kappa(\phi-\phi_{0})\r]\r).~\label{eq:equation-WWI'}
\end{equation}
This potential is same that has been used in~\cite{HSSS:2016}, it is composed of
$\alpha$-attractor potentials~\cite{alpha_attractor} with different slopes 
appearing in the exponent, allowing a discontinuity in the derivative. 
Since in this case the potential is continuous,
\beq
V_0\left(1-\exp[-\alpha\kappa\phi_{\rm T}]\right)=V_1 \left(1-
\exp[-\alpha\kappa(\phi_{\rm T}-\phi_0)]\right),
\eeq
and note that $V_1>V_0$. 
$\kappa^2=8\pi G$ and we use the convention where it is equal to 1. 
The parameter $\alpha$, that controls the slope of the potential, is fixed to $\sqrt{2/3}$ to reproduce the Starobinsky's $R+R^2$ inflationary 
model~\cite{Starobinsky:1980te} 
in the Einstein frame which gives a primordial tensor power spectrum with $r\sim 4\times10^{-3}$ for the featureless case. 
Within this treatment we have only two extra parameters compared to featureless case, $\phi_{\rm T}$ (field value at the 
transition) and $\phi_{0}$ (extent of the discontinuity in the derivative of the potential). Here too $\phi$, $\phi_{\rm T}$ and $\phi_0$ are measured in units of the
reduced Planck mass.
The primordial feature from WWI$'$, as has been discussed in the previous paper on WWI, is similar to 
the original Starobinsky-1992 model~\cite{S92}. 

In both the potentials given in Eq.~\ref{eq:equation-WWI} and~\ref{eq:equation-WWI'}, the tensor-to-scalar ratios are chosen to be 
in perfect agreement with the current upper bounds from the data but also in the ballpark of a possible detection by a CORE like survey. 
\begin{figure*}[!htb]
\begin{center} 
\resizebox{210pt}{140pt}{\includegraphics{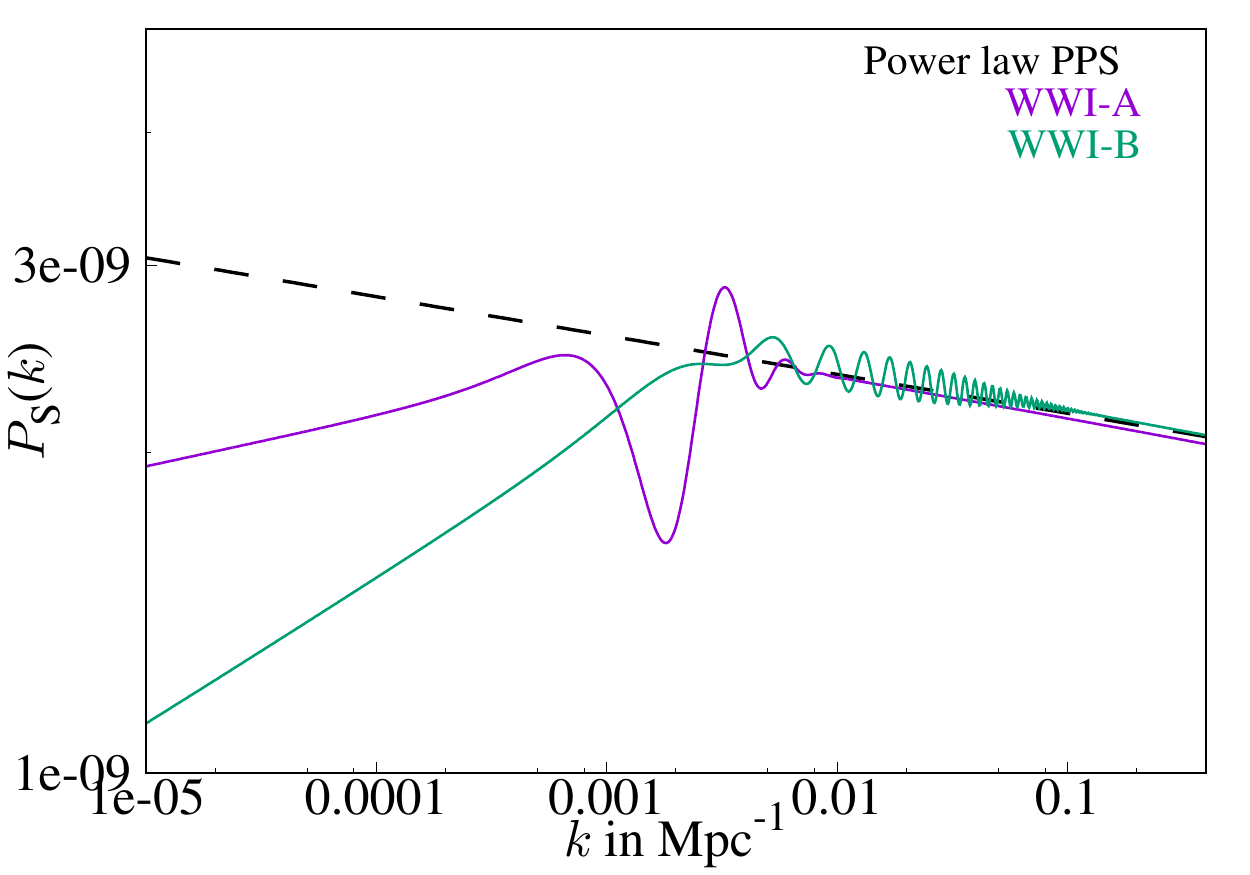}} 
\resizebox{210pt}{140pt}{\includegraphics{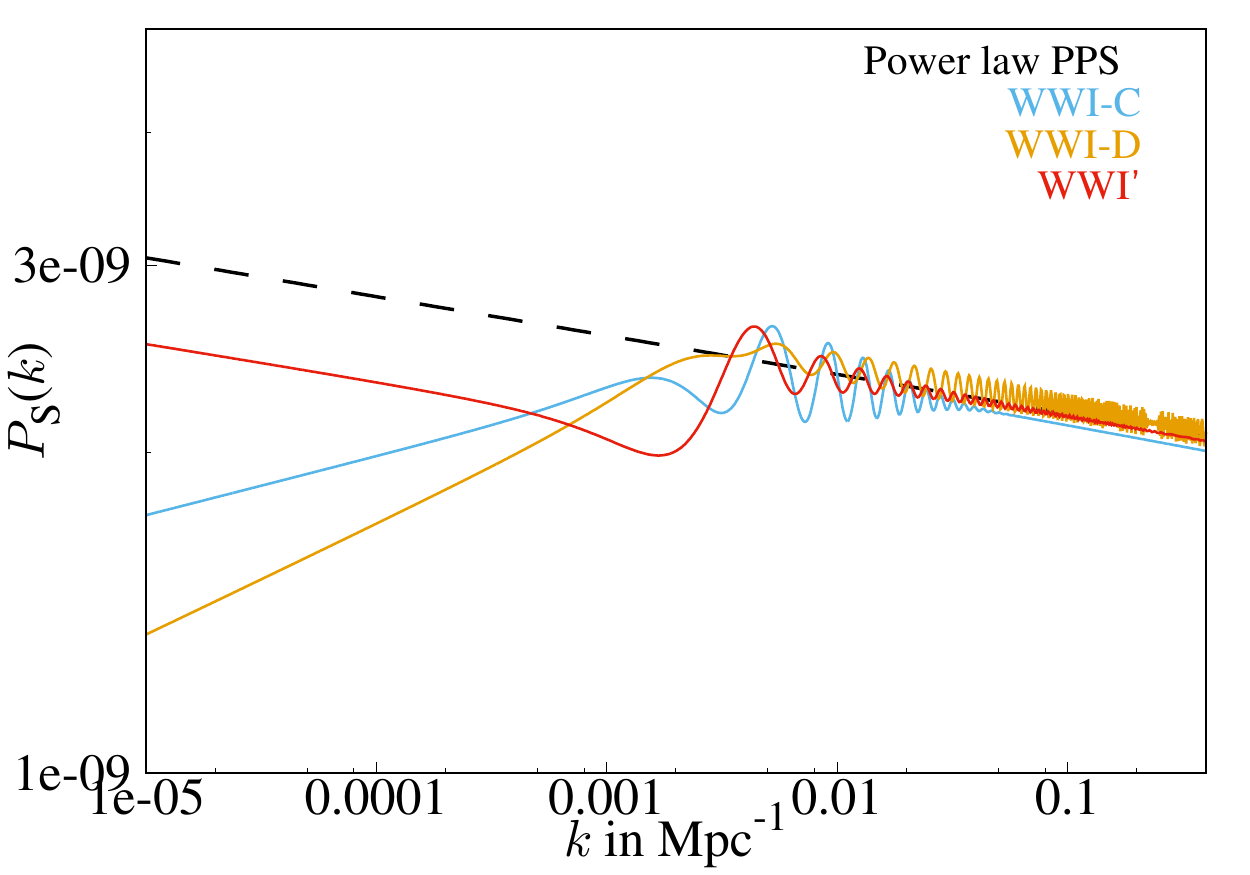}} 
\end{center}
\caption{\footnotesize\label{fig:ppsbestfits} The best fit primordial power spectra obtained from Planck 2015 analysis with Wiggly Whipped Inflation model. Left plot shows WWI-A and B best fits
and the right one shows WWI-C, D and WWI$'$ best fits. Note the best fits WWI-[A, B, C, D] belong to the same potential. Best fit power law PPS is also provided in dashed black. The color codes 
for different best fits will be used consistently throughout the paper.}
\end{figure*}
In~\cite{HSSS:2016} we identified four local minima of the Planck 2015 data, 
corresponding to likelihood peaks, where we observed an improvement in the 
fit compared with the power law in temperature/polarization data separately 
and in combination. We can identify these different types of local and global
features as four broad categories: WWI-A, B, C and D. WWI$'$ having only 2 extra 
parameters provided a substantial improvement in fit to the Planck 2015 data. 
These four+one types of features are plotted in figure~\ref{fig:ppsbestfits}. 
With a CORE like survey, we mainly aim to address what is the expected significance of the features, 
and possibly to distinguish among the four WWI types, should one of them represent the true model of the Universe. 
In the case of WWI$'$ we have only one type of feature, therefore we address in this case to what extent we 
can reject the power law PPS with future CMB surveys, should the best fit WWI$'$ from Planck represent the true model of the Universe.

%\section{Methodology}~\label{sec:data}

\section{Forecasts for WWI}~\label{sec:wwiforecast}

\subsection{WWI and WWI\texorpdfstring{$'$}{'} numerical set up}

The numerical details regarding the solution to WWI models have been 
already discussed in earlier papers~\cite{HSSS:2016,Hazra:2014WI,Hazra:2014WWI}.
%concerning WWI~\cite{HSSS:2016,Hazra:2014WI,Hazra:2014WWI}. 
Hence, here we will only briefly mention 
the main codes that are used in this analysis.
Publicly available code BI-spectra and Non-Gaussianity Operator, {\tt BINGO}~\cite{Hazra:2012BINGO,BINGOPAGE} is used to generate 
the power spectra from WWI and WWI$'$ inflationary models. The discontinuity in the potential is modeled with a Tanh step function with a width $\delta$ as explained before. 
We use the same initial values as has been used in Ref~\cite{HSSS:2016}. {\tt BINGO} solves the background and Mukhanov-Sasaki equations using standard Bunch-Davies 
initial conditions and the initial value of scale factor is estimated by imposing that the $k=0.05~{\rm Mpc}^{-1}$ mode leaves the Hubble radius 50 {\it e-folds} prior to the end of inflation.

We use publicly available {\tt CAMB} and 
{\tt COSMOMC} for our analysis. The codes are modified in order to incorporate the WWI through {\tt BINGO} and the new reionization history. 
In standard Tanh reionization scenario, we use the baryon density ($\Omega_{\rm b} h^2$), cold dark matter density ($\Omega_{\rm CDM} h^2$), the ratio of the sound horizon to the angular diameter distance at decoupling 
($\theta$) and the optical depth ($\tau$) as free parameters. Note that the standard {\it nearly instantaneous} reionization history is 
described by one parameter $\tau$. These four parameters will be common to all the analyses in this paper. When using real data from Planck 2015 release we will allow also the 
foreground and nuisance parameters to vary according to the Planck likelihood set up.
Extra parameters corresponding to inflationary models and reionization used in this paper will be mentioned in the subsequent sections.

\subsection{Priors for WWI and WWI\texorpdfstring{$'$}{'}}
$V_0$ mainly dictates the amplitude of scalar 
perturbations and we have chosen a prior wide enough to ensure the convergence of the likelihood 
bounded from both ends. The field value corresponding 
to when the phase transition in inflation occurs, $\phi_{\rm T}$, is varied such that the
features occur on the cosmological scales probed by an experiment like CORE. 
The potential parameter $\gamma$ has priors that include no 
suppression (for $\gamma=0$) and a value where the transition allows enough {\it e-folds} ($\sim 70$) of inflation.
Similarly, for $\phi_{0}$, that represents the amplitude of the 
wiggles in the PPS, we set the lower limit to zero, which corresponds to no wiggles. The higher limit is 
$\phi_{0}=0.04$ $M_{\rm Pl}$
that shows oscillations (for all widths of transition considered) that are large enough to be strongly disfavored by Planck 2015 data. 
$\ln(\delta)$ is varied from -12 to -3 that allow sufficiently sharp and wide transitions.
For the WWI we calculate the angular power spectrum at all multipoles instead of 
interpolating, like usually done in numerical codes, in order to capture the sharp features.

%{\color{blue}devo piazzarla da qualche parte ma non qui for Reionization part, where we only use power law 
%as the primordial power spectra, we use interpolation.}

%\section{Forecasts for CORE}

\subsection{CORE specifications}

We now describe the methodology we use 
to forecast the capabilities of a concept for a future CMB space mission as CORE to constrain the primordial origin of the features. 
As fiducial cosmologies, we choose the WWI best-fits of the Planck 2015 temperature and polarization data~\cite{HSSS:2016}.
We perform the CORE forecasts as described in~\cite{core:inf,core:cosmoparam}. 
We use an inverse Wishart likelihood with an effective noise sensitivity and angular resolution obtained from an inverse noise weighted 
combination of the central frequency CORE channels whose specifications are:
\begin{eqnarray}
 {\mathrm{Frequency\, [GHz]}}&=& \Big\{130,145,160,175,195,220\Big\}\nonumber\\
 \mathrm{FWHM\, [Arcmin]}&=&\Big\{8.51,7.68,7.01,6.45,5.84,5.23 \Big\}\nonumber \\
 \mathrm{\Delta T\, [\mu K\,arcmin]}&=&\Big\{3.9,3.6,3.7,3.6,3.5,3.8 \Big\}\nonumber \\
 \mathrm{\Delta P\, [\mu K\,arcmin]}&=&\Big\{5.5,5.1,5.2,5.1,4.9,5.4 \Big\}\nonumber \\
\end{eqnarray}
We assume that systematic effects are subdominant and that foreground contaminations are kept under control 
by the lower 6 (down to 60 GHz) 
and higher 7 (up to 600 GHz) frequency channels \cite{core:inst,Delabrouille:2017rct}. 
In addition to temperature and polarization anisotropies, we consider the CMB lensing potential 
power spectrum $C_\ell^{\rm PP}$, which an experiment like CORE will be able to reconstruct up to the scales where linear theory is reliable. 
As simulated noise spectrum in the lensing potential we use the EB estimator~\cite{Hu:2003vp} as in~\cite{Errard:2015cxa,core:cosmoparam,core:inf}. 

Since we are investigating the possible constraints on features beyond the standard model in the primordial power 
spectrum with CORE, for comparison we present the 95\% CL uncertainties expected from an experiment like CORE compared to 
the Planck 2015 results in the standard $\Lambda$CDM power law PPS: 
\begin{eqnarray}
{\mathrm {Parameter}}&=&\{\Omega_{\rm b}h^2,\Omega_{\rm CDM}h^2,\tau,H_0,\ln[10^{10} A_{\rm S}],n_{\rm S},\sigma_8\};\nonumber\\
\sigma_{\mathrm {Planck}}&=&\{0.0003,0.0029,0.034,1.29,0.066,0.01,0.026\};\nonumber\\
\sigma_{\mathrm {CORE}}&=&\{0.00007,0.0006,0.004,0.23,0.008,0.003,0.002\};~\label{Eq:PlanckCoreArray}
\end{eqnarray}
As already noticed in~\cite{core:cosmoparam,core:inf}, there is an improvement of almost an order of magnitude on the standard cosmological 
model constraints given by CORE with respect to Planck 2015. The improvement is mainly due 
to the better sensitivity in polarization and lensing measurements for CORE, although we need to bear in mind the differences between a real Planck 
likelihood including marginalization on nuisance parameters and the simplified ideal CORE likelihood where
no residual foreground or systematic uncertainties, outside the instrumental error, are taken into account.
%anyway we note that the Planck likelihood, which is performed on real data with two 
%different numerical algorithms, also contains the marginalization on the foreground residual contamination which may lead to a slighlty 
%larger errors with respect  to a simulated perfectly clean mock dataset  like the CORE-like we are considering.
%parameters~\cite{core:cosmoparam,core:inf}. 
Above and in the following, we also present the constraints for the 
Hubble parameter ($H_0$) today and $\sigma_8$ (in some of the cases), which are derived parameters. All the errors presented here represent 2$\sigma$ confidence intervals. 
%Such comparison for power law and extended power law model has been performed in `Exploring Cosmic Origins with CORE: Inflation'~\cite{core:inf}.   

%\subsection{Constraints on the primordial features}

\subsection{Forecasts for primordial features}

%We already presented the results for the WWI corresponding to Planck 2015 data, although  in a slightly different 
%form, in~\cite{HSSS:2016}. 
The results for the WWI models with Planck 2015 data, although in a slightly different
form, were presented in~\cite{HSSS:2016}.
In the following results we use the best fits of~\cite{HSSS:2016} as fiducial values for the CORE simulated data. 
Though in this paper we use a slightly modified form with respect to~\cite{HSSS:2016}, note that since the latter is nested 
in the more general form used here the local best fits will not change; a simple change of variable translates the old best 
fit parameters to the modified parametrization.

%%%%%%%%%%%%%%%%%%%%%%%%%%%%%%%%%%%%%%%%%%%%%%%%%%%%%%%%%%%%%%%%%%%%%%%%%%%%%%%%%%%%%%%%%%%%%%%%%%%%%%%%%%%%%%%%%%%%%
\renewcommand{\arraystretch}{1.1}
\begin{table*}[!htb]
\begin{center}
\vspace{4pt}
\begin{tabular}{| c | c | c |c |c|c|}
\hline\hline
\multicolumn{6}{|c|}{\bf Parameter Constraints for WWI}\\
\hline
Parameters & WWI-A& WWI-B & WWI-C&WWI-D&WWI$'$\\

&$\Delta_{\rm DOF}=4$&$\Delta_{\rm DOF}=4$&$\Delta_{\rm DOF}=4$& $\Delta_{\rm DOF}=4$& $\Delta_{\rm DOF}=2$ \\
\hline
 $\Omega_{\rm b}h^2$ 
 &$2.217\pm0.006$
 &$2.222\pm0.006$
 &$2.217\pm0.006$
 &$2.221\pm0.006$
 &$2.218\pm0.006$\\
 $\times10^{2}$& & & & &\\
\hline
 $\Omega_{\rm CDM}h^2$ 
 &$12.05\pm0.06$
 &$12.03\pm0.06$ 
 &$12.06\pm0.06$ 
 &$12.04\pm0.06$ 
 &$12.03\pm0.06$\\
 $\times10^{2}$& & & & &\\
 \hline
 $\tau$ 
 &$0.081\pm0.005$
 &$0.093\pm0.005$ 
 &$0.074^{+0.005}_{-0.004}$ 
 &$0.092\pm0.005$ 
 &$0.085\pm0.005$  \\
\hline
 $H_0$ 
 &$66.93\pm0.25$
 &$67.06\pm0.24$ 
 &$66.9^{+0.24}_{-0.23}$ 
 &$67.03^{+0.25}_{-0.24}$  
 &$67.02\pm0.23$ \\
\hline
\hline
 $\ln[10^{10} V_{0}]$ 
 &$1.735\pm0.009$
 &$1.754\pm0.009$ 
 &$1.72^{+0.009}_{-0.008}$ 
 &$1.757\pm0.009$ 
 &$0.2809\pm0.0009$ \\
\hline
$\phi_{0}$ 
&$0.013^{+0.008}_{-0.007}$
&$0.0037^{+0.0019}_{-0.0021}$
&$0.0057\pm0.0022$ 
&$0.003\pm0.0005$  
&$0.1\pm0.06$\\
\hline
 $\gamma$ 
 &$0.03^{+0.04}_{<}$
 &$0.041^{+0.035}_{<(-0.02)}$
 &$0.026^{+0.028}_{<(-0.02)}$  
 &$0.034^{+0.02}_{-0.03}$
 &-\\
 \hline
 $\phi_{\rm T}$ 
 &$7.887^{+0.012}_{-0.014}$
 &$7.9104^{+0.0012}_{-0.0014}$
 &$7.91^{+0.0016}_{-0.0017}$
 &$7.9111\pm0.0005$ 
 &$4.508^{+0.003}_{-0.002}$\\
\hline
 $\ln[\delta]$ 
 &$-4.4^{+0.7}_{-0.8}$
 &$-6.89^{+2.2}_{-1.3}$
 &$-5.91^{+0.86}_{-0.73}$ 
 &$10.2^{+1.6}_{<}$ 
 &- \\
\hline\hline
\end{tabular}
\end{center}
\caption{~\label{tab:bounds2}Background and inflationary parameter constraints WWI projected for CORE. For WWI potential we present the constraints on the parameters when CORE data is simulated using the 
four local best fits to the Planck data. For WWI$'$ potential also we provide the constraints expected from CORE. Errors correspond to 95\% confidence. For parameters that are unbounded at $2\sigma$ 
within the priors provided are denoted as $<$ (unbounded from below). If a value is quoted after $<$ in parenthesis, it implies the 68.3\% error in that direction.
Absence of error in parenthesis following $<$ denotes that the parameter is unbounded even in $1\sigma$ in that direction. The extra degrees of freedom compared to the strict slow-roll 
part inflation are provided by $\Delta_{\rm DOF}$.} 
\end{table*}
%%%%%%%%%%%%%%%%%%%%%%%%%%%%%%%%%%%%%%%%%%%%%%%%%%%%%%%%%%%%%%%%%%%%%%%%%%%%%%%%%%%%%%%%%%%%%%%%%%%%%%%%%%%%%%%%%%%%%

%%%%%%%%%%%%%%%%%%%%%%%%%%%%%%%%%%%%%%%%%%%%%%%%%%%%%%%%%%%%%%%%%%%%%%%%%%%%%%%%%%%%%%%%%%%%%%%%%%%%%%%%%%%%%%%%%%%%%
%%%%%%%%%%%%%%%%%%%%%%%%%%%%%%%%%%%%%%%%%%%%%%%%%%%%%%%%%%%%%%%%%%%%%%%%%%%%%%%%%%%%%%%%%%%%%%%%%%%%%%%%%%%%%%%%%%%%%
%%%%%%%%%%%%%%%%%%%%%%%%%%%%%%%%%%%%%%%%%%%%%%%%%%%%%%%%%%%%%%%%%%%%%%%%%%%%%%%%%%%%%%%%%%%%%%%%%%%%%%%%%%%%%%%%%%%%%
\begin{figure*}[!htb]
\begin{center} 
\resizebox{220pt}{160pt}{\includegraphics{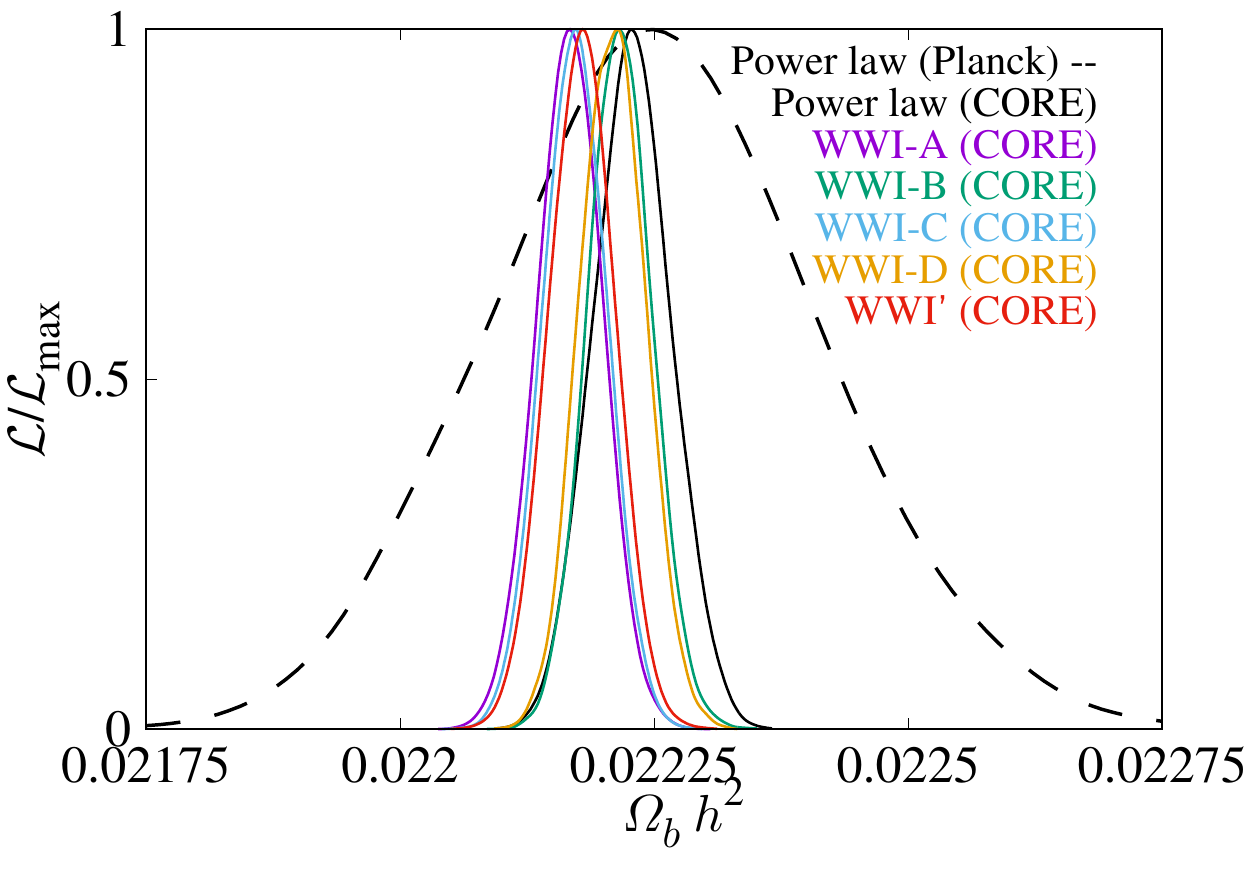}} 
\hskip -8pt\resizebox{220pt}{160pt}{\includegraphics{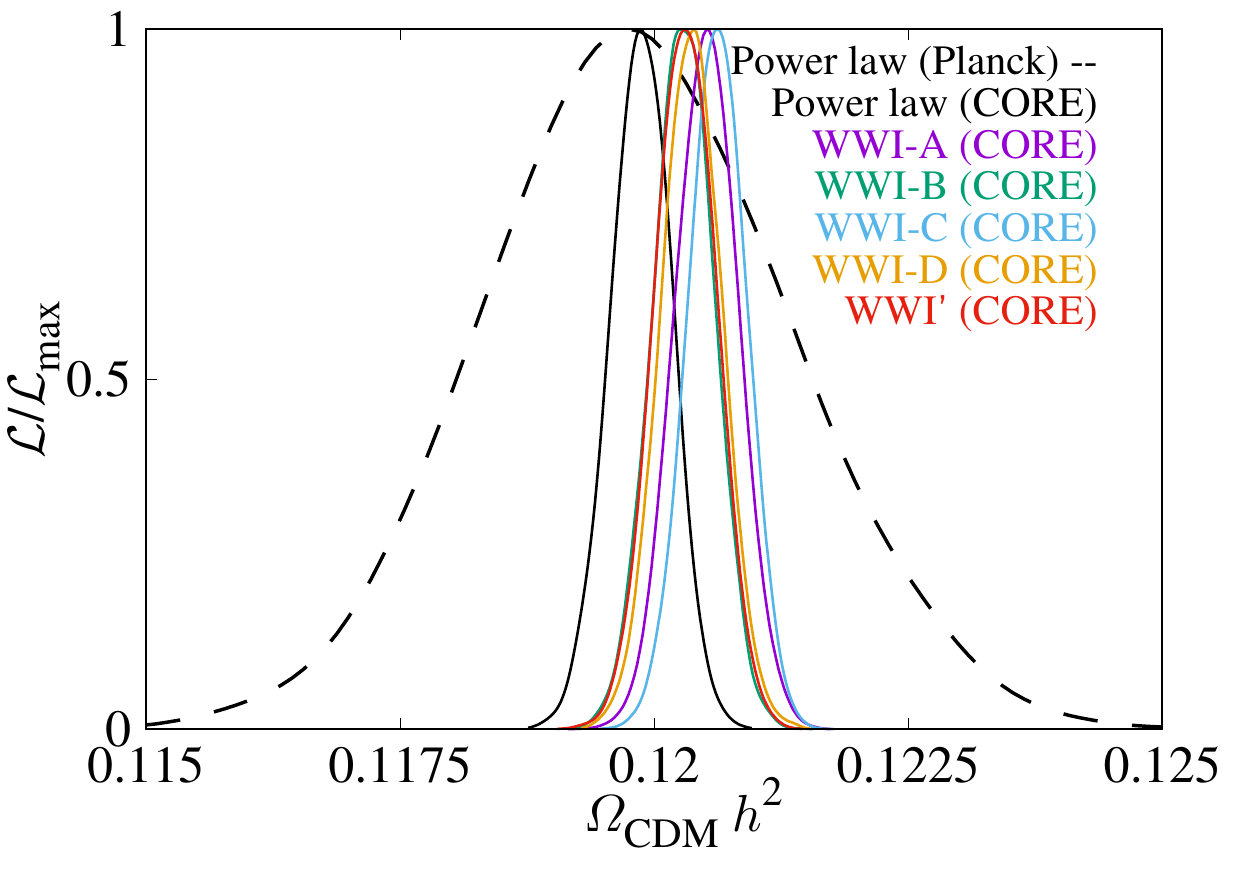}} 

\resizebox{220pt}{160pt}{\includegraphics{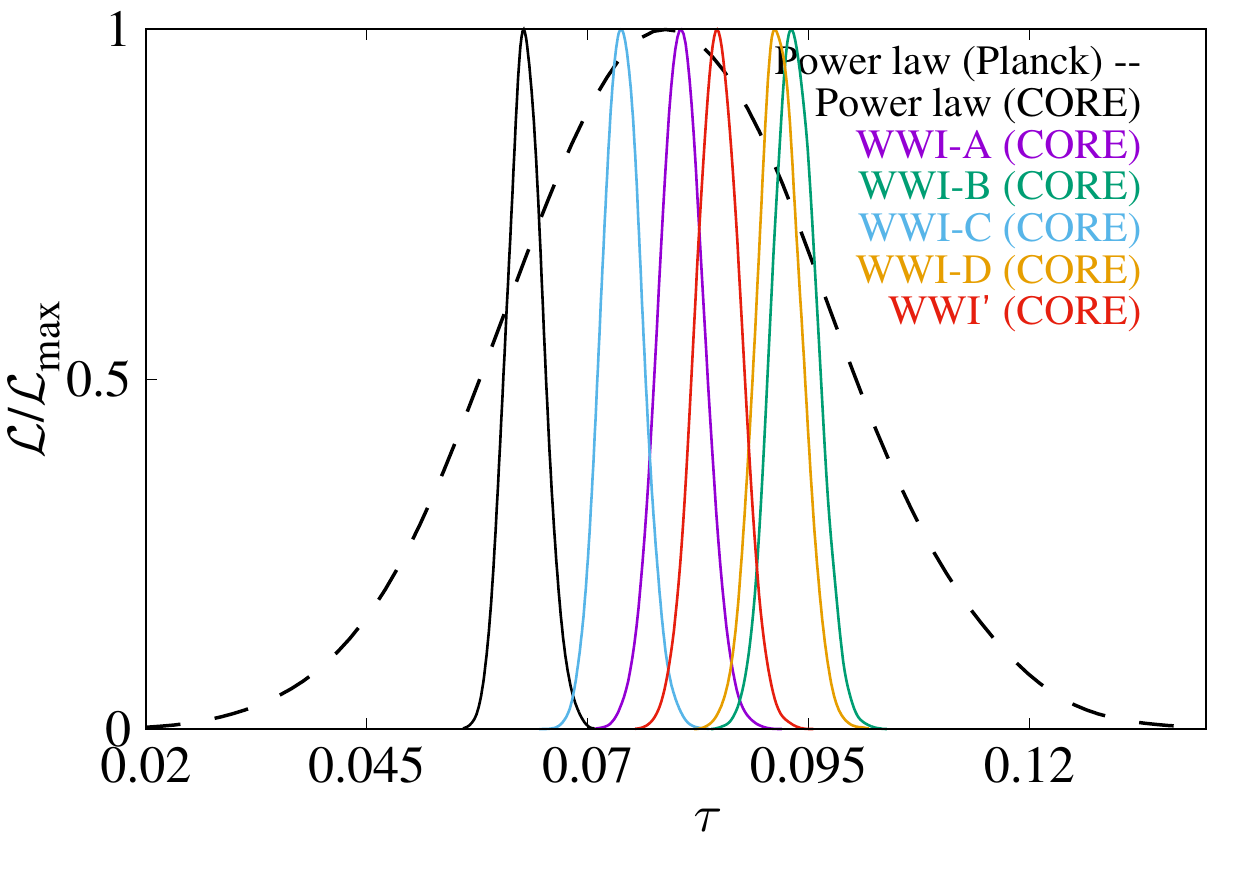}} 
\hskip -8pt\resizebox{220pt}{160pt}{\includegraphics{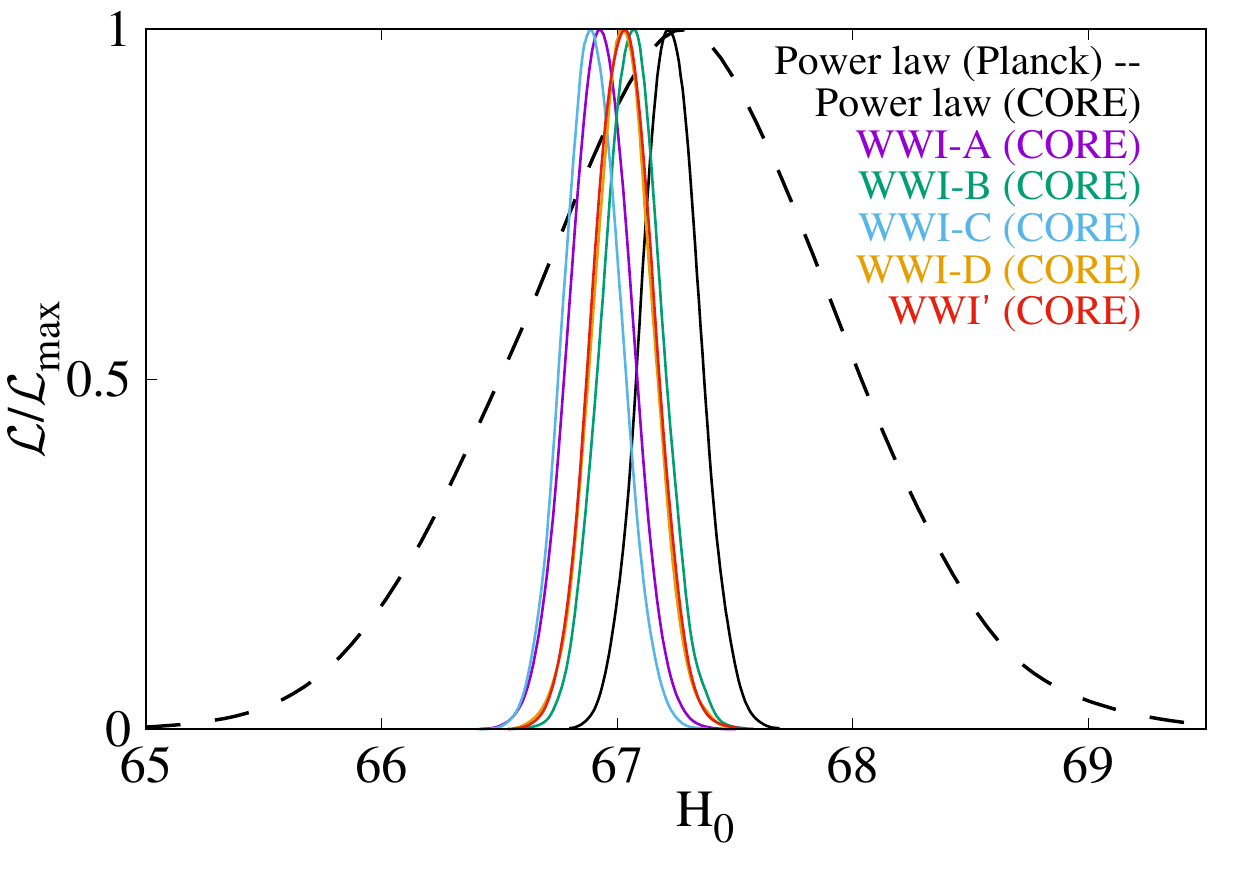}} 
\end{center}
\caption{\footnotesize\label{fig:bgnd-param}Background cosmological parameter constraints for power law and the WWI. For the power law, constraints obtained using Planck 
data are provided in dashed black and projected CORE constraints are provided in solid black. For the other cases, we just provide the results from CORE in solid line. A couple 
of points to note in these plots are: firstly, as expected, we obtain significant improvement in the constraints if we use simulated CORE sensitivity; secondly since each of the 
WWI local best fits from Planck is used for the CORE simulation as fiducial model, we find the likelihoods localized at different region in the background parameter space and for 
certain parameters, there are very little overlap between the likelihoods, indicating the ability to distinguish them in the future CMB observations.}
\end{figure*}

\begin{figure*}[!t]

\begin{center} 
\resizebox{175pt}{110pt}{\includegraphics{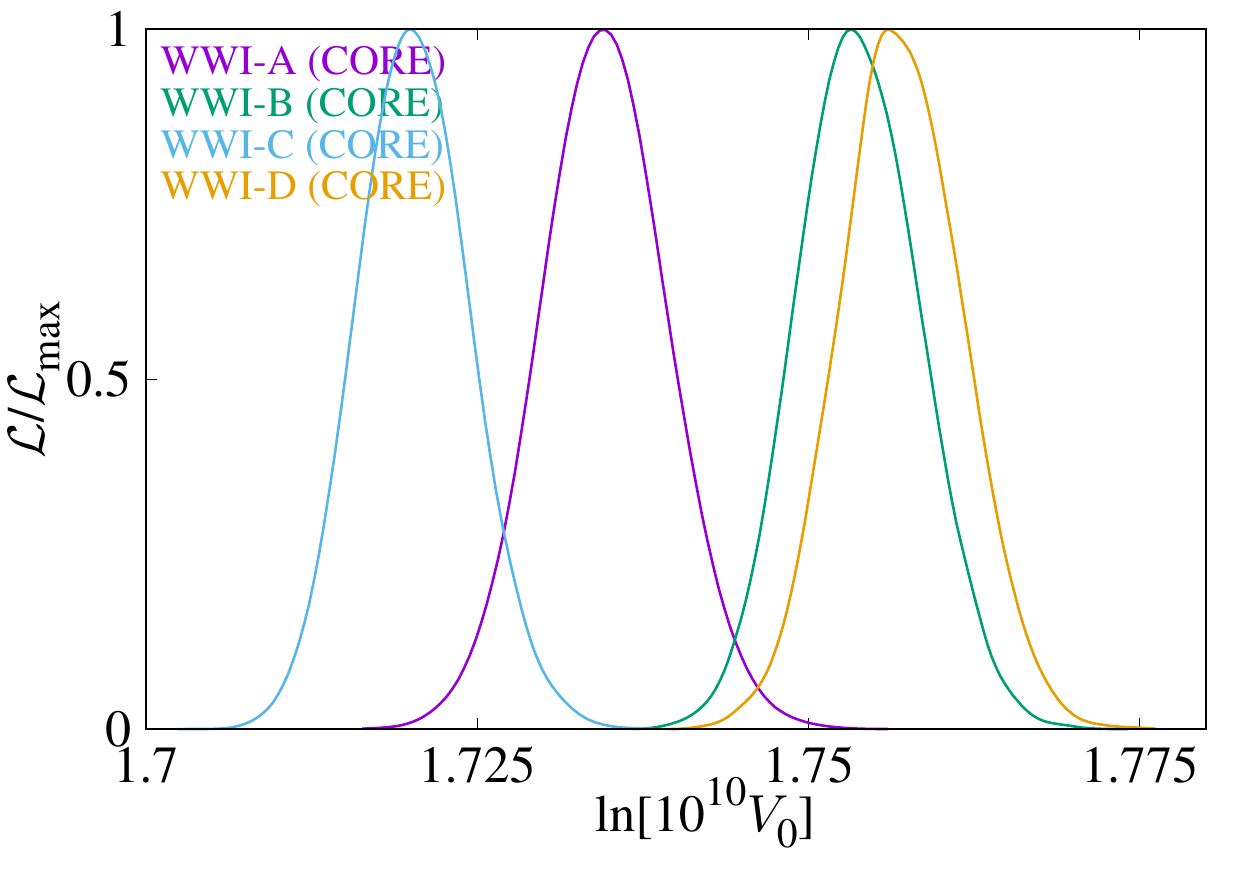}} 
\resizebox{175pt}{115pt}{\includegraphics{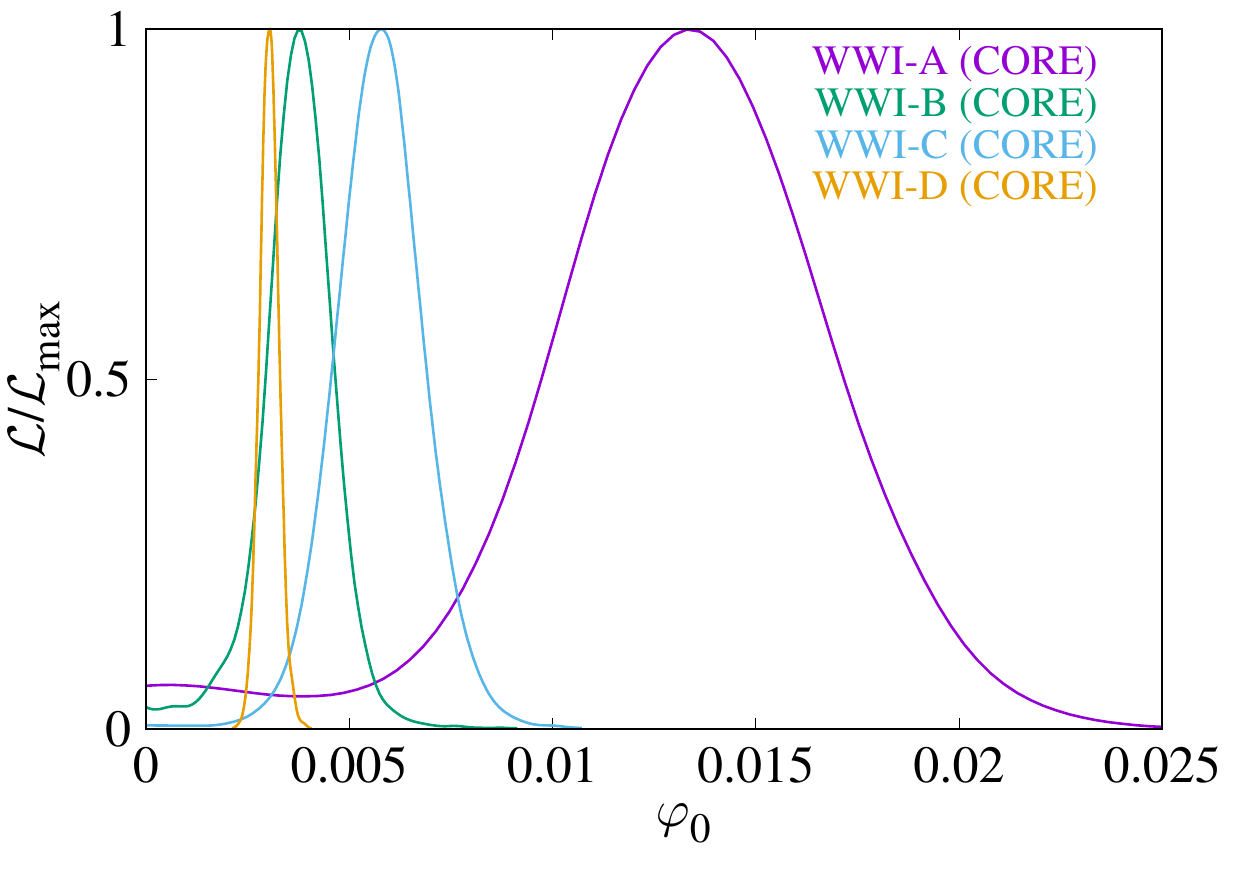}} 

\resizebox{170pt}{115pt}{\includegraphics{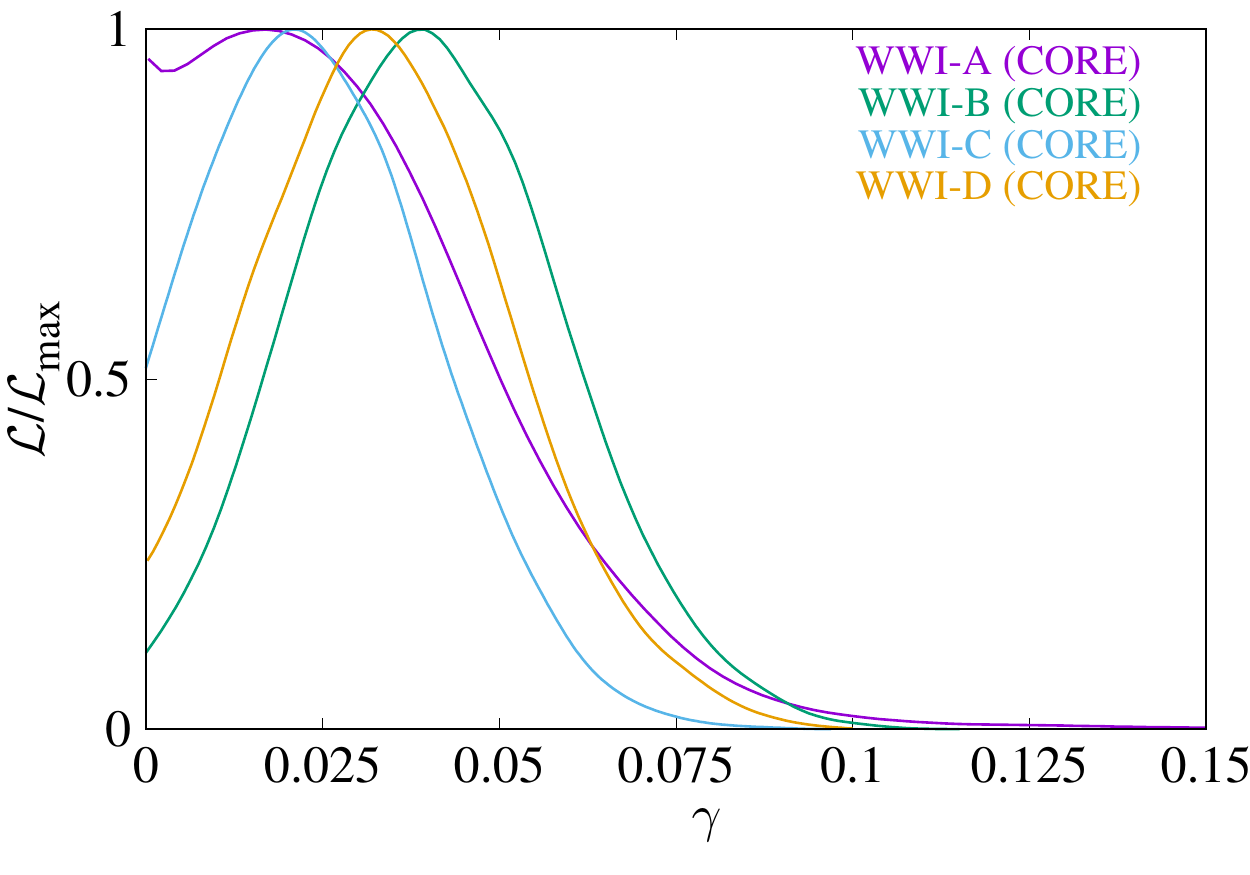}} 
\resizebox{170pt}{115pt}{\includegraphics{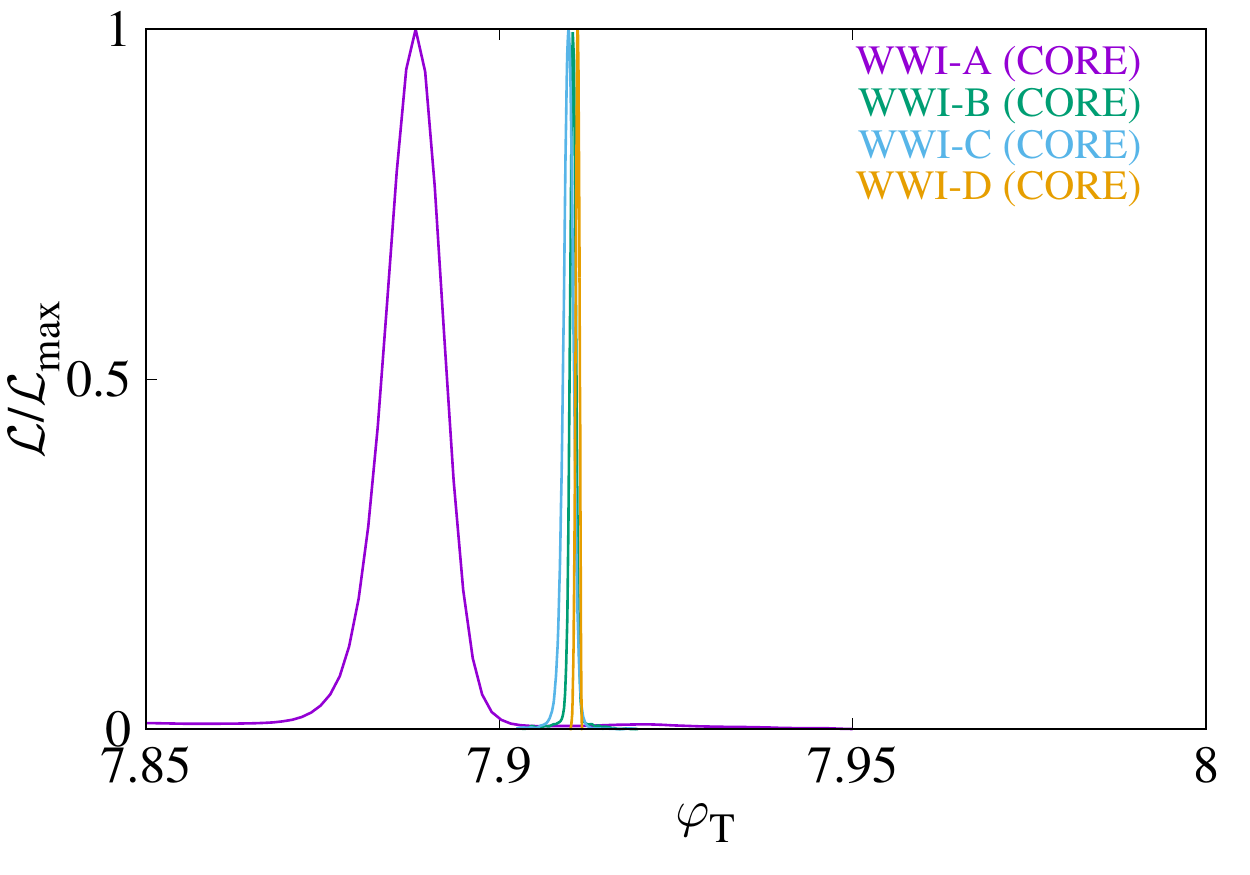}} 

\resizebox{170pt}{115pt}{\includegraphics{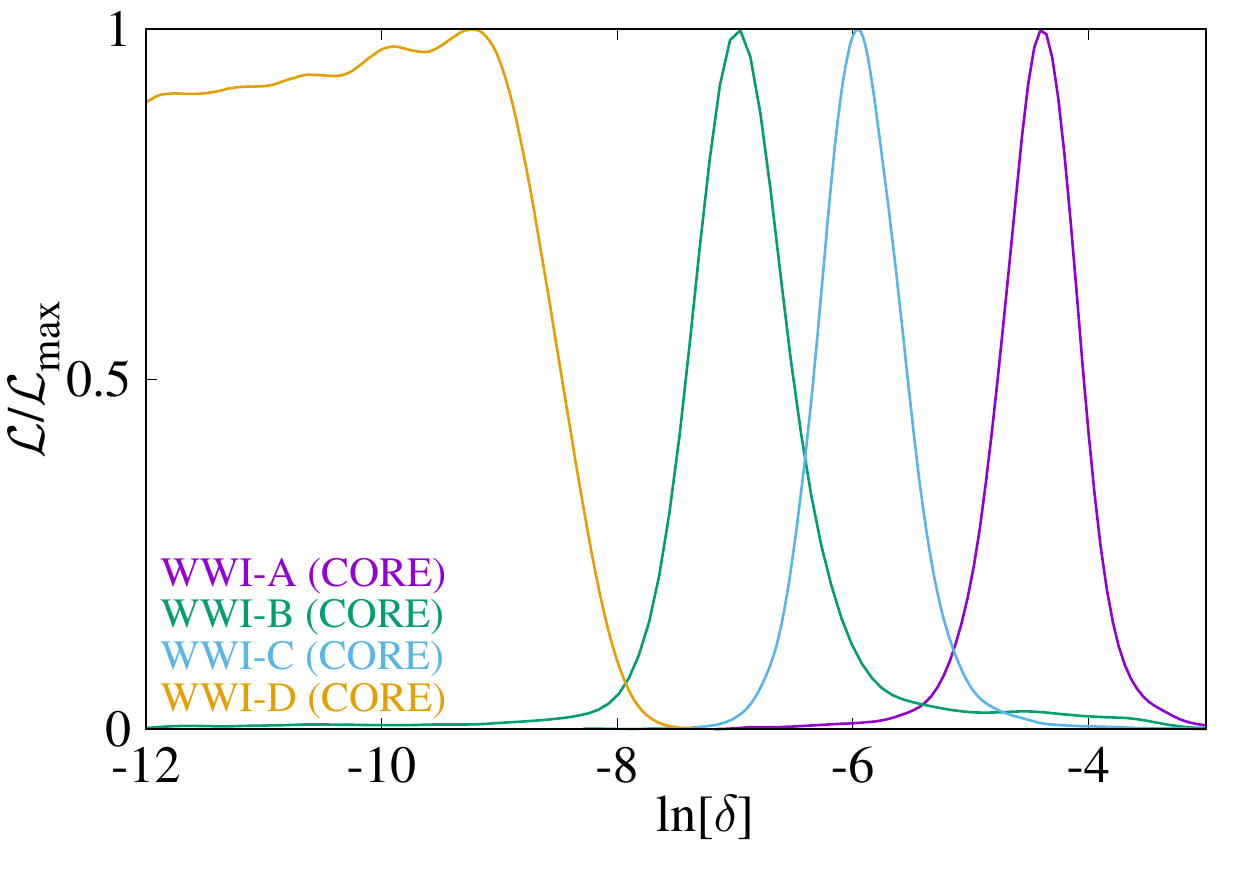}} 
\resizebox{170pt}{115pt}{\includegraphics{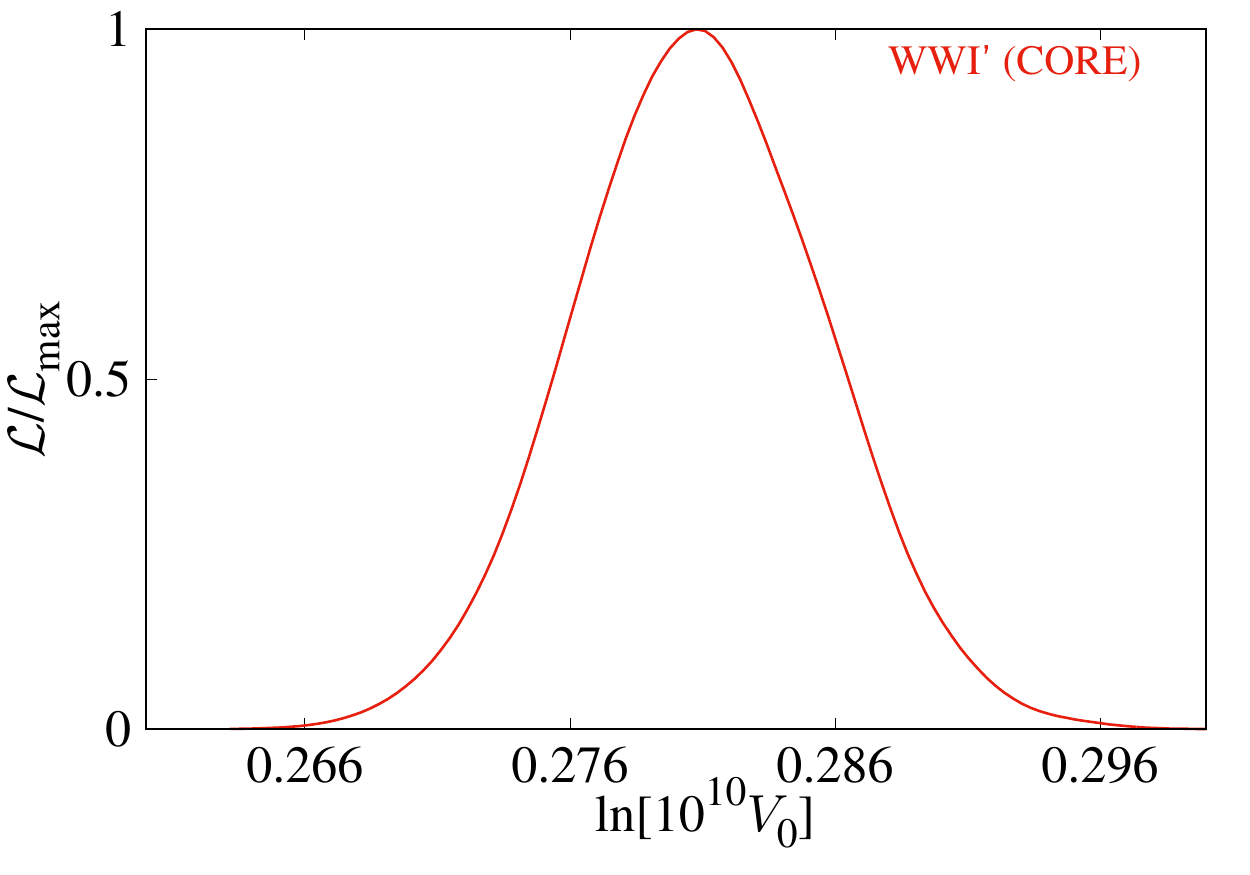}} 

\resizebox{170pt}{115pt}{\includegraphics{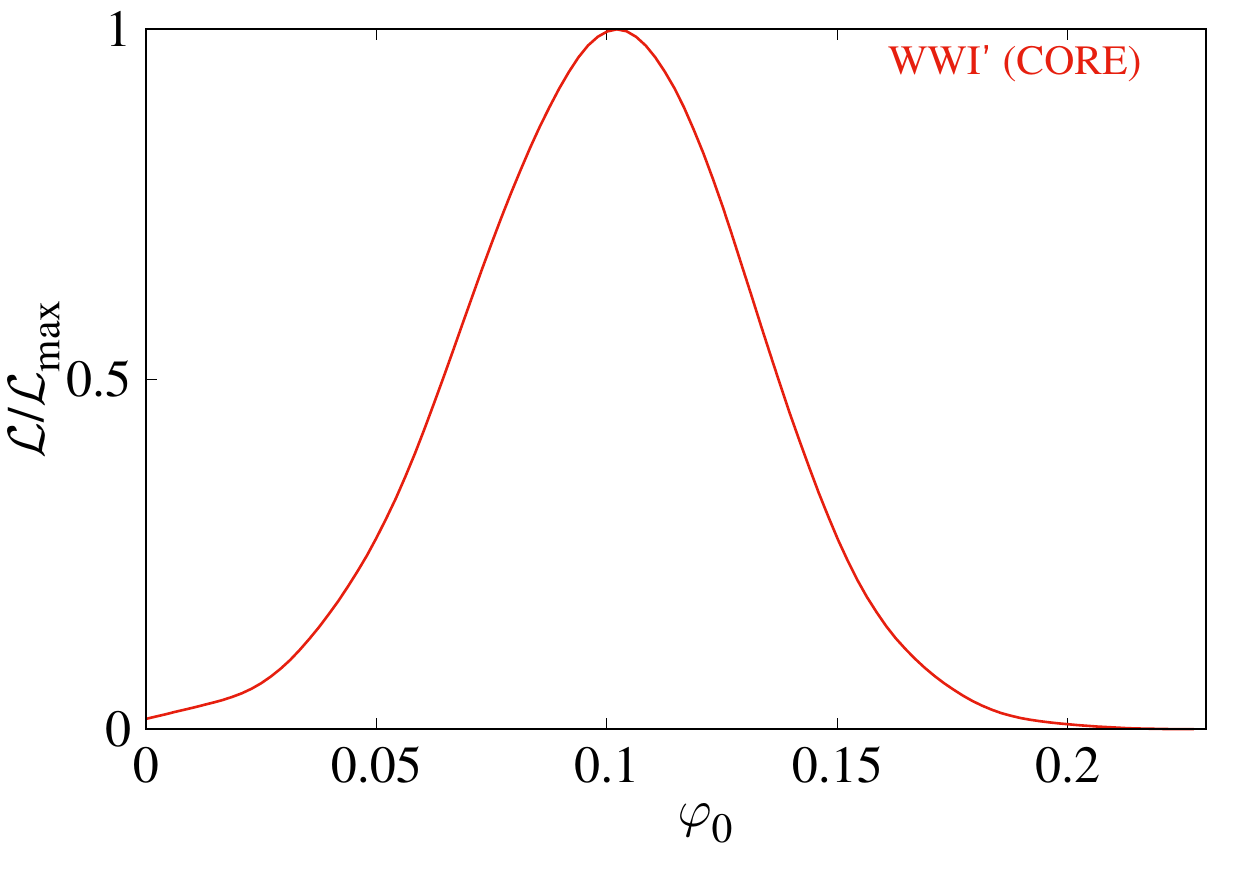}} 
\resizebox{170pt}{115pt}{\includegraphics{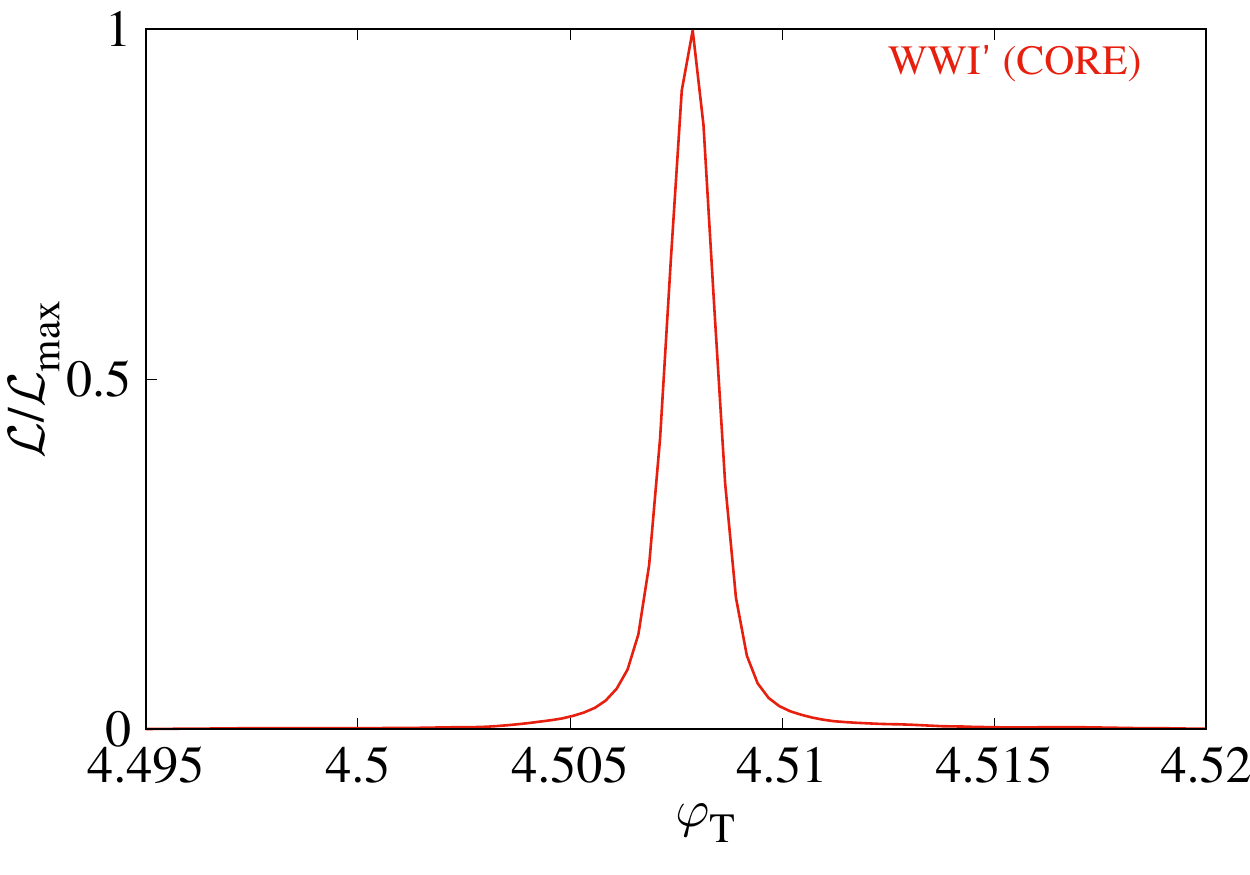}} 
\end{center}
\caption{\footnotesize\label{fig:inf-param}Constraints on inflationary potential parameters for WWI-[A,B,C,D] and WWI$'$. The first 5 plots from top with different posterior distributions correspond to WWI-[A,B,C,D]
features and the last three single likelihoods represent WWI$'$. Note that for WWI potential, while Planck data can indicate different types of features as its local minima, these sets of plots
are clearly showing that some of the features, should it represent the true model of the Universe, will be significantly detected with CORE. Wiggles in the WWI are characterized by $\phi_{0}$
hence, the distance of $\phi_{0}=0$ from the maximum likelihood will dictate the significance of that type of wiggles with future data. The large scale suppression (generated by $\gamma$) 
will not be constrained well by future data owing to the cosmic variance. The amplitude of perturbations (controlled principally by $V_0$), however is constrained with high 
significance for all the models. Features continuing to small scales are found to have higher chances of being detected.}
\end{figure*}
%%%%%%%%%%%%%%%%%%%%%%%%%%%%%%%%%%%%%%%%%%%%%%%%%%%%%%%%%%%%%%%%%%%%%%%%%%%%%%%%%%%%%%%%%%%%%%%%%%%%%%%%%%%%%%%%%%%%%
In table~\ref{tab:bounds2}, we list the constraints for CORE in the WWI case. 
We stress again that within the WWI potential, A, B, C, D represent the best fits that we obtained 
using Planck 2015 data in~\cite{HSSS:2016}. Since our objective is to 
predict at which significance level CORE data can constrain these favored models, 
should they be the real underlying model of the Universe, we have used the 
local best fits to the Planck TTTEEE + lowTEB + BKP datasets~\footnote{BKP represents joint likelihood from BICEP2-Keck array that includes dust polarization
measurements from Planck~\cite{Array:2015xqh}. We used the best fit parameters for this likelihood combination because of completeness. However, note that 
in this paper we are not using CORE BB likelihood.}. 
For the WWI$'$ potential it generates one single class of 
features with only two parameters, this class improves the 
fit to both Planck temperature and polarization data compared to power law PPS. 
In the different rows of the table~\ref{tab:bounds2} we provide the background 
and inflationary parameter constraints, the columns represent 
the cases where WWI-A, B, C, D and WWI$'$ best fits are used as fiducial cosmologies for our simulations.
%the CORE-like simulated data. 

Comparing the uncertainties in table~\ref{tab:bounds2} with those in~\ref{Eq:PlanckCoreArray} 
we note that though the PPS is characterized by more parameters, 
the WWI framework seems to constrain the baryon density comparably to the power law case. 
However, as has been discussed, in the WWI framework, we set the 
slow roll part of the potential (apart from the amplitude parameter, $V_0$) 
in a way that the spectral tilt generated in the featureless case is $\sim 0.96$, 
in good agreement with Planck data. We are therefore reducing the space dimensionality 
%the degeneracy 
of the standard cosmological parameters compared to the power law case. 

It is however interesting to note that the presence of the features do not loosen the bounds on the cosmological parameters, except for $\tau$. The reason for this 
is that WWI provides a power suppression on large scales for all its variants, hence the optical depths obtained in these models are 
all substantially different from the power law case, affecting the simulated CORE power spectra. Due to this degeneracy with the PPS shape, the constraints expected from 
CORE are slightly degraded. 

In figure~\ref{fig:bgnd-param} we show the comparison of background parameters from power law and WWI. In order to highlight the present 
constraints, we plot also the Planck constraints on the concordance $\Lambda$CDM model in dashed black.

%  In these four sets of plots it is important to note how different features 
%prefer different cosmological parameters. While they all fall within the uncertainties of the Planck power law constraints, with the power of a CORE-like dataset 
%we can distinguish among the different cosmological parameters associated with different features
We stress that different features provide different best fits to the Planck 2015 
and therefore in the CORE simulated data have different fiducial values for the cosmological parameters.
The results show that while 
with Planck data all the different background parameter best fits 
fall within the uncertainties of the power law constraints, making it impossible to distinguish one from the other, 
with CORE it is possible not only to constrain some of the parameters 
which characterize the features but also distinguish among the different 
cosmological parameters associated with different features themselves increasing the possibility to 
distinguish among the possible models.
For example, for baryon and CDM densities, 
the posterior peaks for WWI-A and C are quite off 
with respect to the power law. Most evidently we have differences for
the optical depth, in fact, as discussed before, 
we find that $\tau$ from all the feature models 
can be distinguished. For the Hubble parameter 
we note a common trend as has been discussed in earlier 
papers~\cite{Hlozek:2011pc,Blanchard:2003du,Hunt:2007dn,HSS:freeform}. 
The presence of any kinds of features, within the flexibilities of WWI framework, prefers a lower value of $H_0$ compared to power law.

Concerning the inflaton potential constraints we have five parameters for 
the WWI potential (four of which are the extra parameters to describe the beyond slow roll part) and three 
for WWI$'$ potential (two extra parameter with respect to the canonical $\alpha$-attractor case).
Again all the bounds in table~\ref{tab:bounds2} are the 95\% C.L. 
These constraints are extremely important in order to predict 
the probability of a specific feature model being detected in the future. Some parameters for the WWI, 
in particular $\gamma$ and $\delta$, remain partially unconstrained. 
The marginalized posteriors corresponding to these inflationary potential parameters are plotted in figure~\ref{fig:inf-param}.

We note how the scale of inflation $V_0$ will be tightly 
constrained by an instrument like CORE, as it fixes the overall CMB normalization scale. 
The parameter $\phi_{0}$ denotes the amplitude of primordial 
oscillations as it determines the height of discontinuities in the potential (for WWI potential)
or in its derivative (for WWI$'$ potential). Hence, the greater is 
the distance of $\phi_{0}=0$ from the peak of the posterior the higher is the confidence to rule out the power law model, 
assuming that the corresponding inflationary feature represents the true model of the Universe. 
The results show how with an experiment like CORE it would be 
possible to detect the features chosen within the WWI framework, 
from moderate to high significance, should they represent the real Universe~\footnote{We define moderate significance by $<3\sigma$ C.L.}.
For example WWI-A and WWI-B features are constrained to 2-3$\sigma$ level, WWI-C and WWI$'$ have $\sim 3\sigma$ significance.
WWI-D, where oscillations extend to small scales (all the scales probed by CORE), has the highest probability of 
being detected ($\gg3\sigma$) due to its small scale signature which is in the optimal range of CORE. 

The extent of large scale suppression in these models, controlled by $\gamma$, however, is not strongly constrained in any of the features 
from WWI potential, even with a sensitivity and a sky coverage of an instrument like CORE, mainly due to cosmic variance. 
Of these suppression, WWI-B which has the largest suppression amongst all the PPS, obviously has the best probability ($\sim2\sigma$)
as can be noted from table~\ref{tab:bounds2}.

The parameter $\phi_{\rm T}$ is related to the location of the feature. Its constraints indicate at which scale particular feature is 
favored by the data and that translates exactly to the field value of the phase transition during inflation. 
Note that the feature positions in all the cases are well constrained. However, the variance in the position from WWI-A is largest as 
it represents the largest scale feature and a small shift in the feature position do not significantly degrade the likelihood.

The width of the step, that defines the sharpness of the wiggles, is also an indicator for the evidence of the features. 
Different types of features are classified also by the characteristic frequency of the oscillations, the constraints on the
$\ln[\delta]$ in WWI potential show posterior distributions with peaks localized in four different regions in the parameter space opening the possibility to 
distinguish among different models also through this 
parameter. The width in WWI-A, B, C and WWI$'$ are well constrained whereas for WWI-D we have an 
upper bound which shows that steps sharper than $\ln[\delta]\sim-10$ will not be constrained. 
%with CORE-like data. 

\section{Reionization history}~\label{sec:reion}

Einstein-Boltzmann codes, such as {\tt CAMB}~\cite{cambsite,Lewis:1999bs}, typically use a fixed form to model the reionization history, namely it is assumed that the free electron fraction
follows a hyperbolic tangent in the following form: 
\begin{equation}
 x_e(z)=\frac{1+F_{\rm He}}{2}\l[1+\tanh\l(\frac{y(z_{\rm re})-y(z)}{\delta_{\rm Reion}}\r)\r],~\label{eq:reion-tanh}
\end{equation}
where, $y(z)=(1+z)^{3/2}$. ${\delta_{\rm Reion}=1.5\sqrt{1+z_{\rm re}}dz}$ where $dz$ is fixed to be 0.5 in Planck baseline analysis. 
$x_e(z)$ is the free electron fraction {\it w.r.t.} hydrogen, the factor $1+F_{\rm He}$ accounts for the corrections due to the fraction of Helium and $F_{\rm He}$ in this 
treatment is derived consistently with Big Bang Nucleosynthesis and it approximately takes the value 0.08.
In addition, at a redshift of 3.5, considering doubly ionized Helium, another Tanh step for Helium with an asymptotic factor of $F_{\rm He}$
is added to the electron fraction. 
$z_{\rm re}$ denotes the redshift where the free electron fraction from hydrogen is 0.5. Defining the reionization history in this way,
{\tt CAMB} does a search in the $z_{\rm re}$ and find the redshift of reionization 
for a given optical depth $\tau$ which is one of the baseline parameter in the Markov-Chain 
Monte Carlo analysis code {\tt COSMOMC}~\cite{cosmomcsite,Lewis:2002ah}.

%In this parametrization of reionization history, once $\Delta_{\rm Reion}$, the width of the step, is fixed everything 
%is fixed and there is no freedom to play with. 
%In addition, within this model there is the possiblity to have an electron fraction $x_e<1$ at low redshift  $z<5\sim6$ which would be in strong disagreement with current quasar observations which 
%show a fully ionized universe at those redshift~\cite{Fan:2006dp}.
In order to allow for more freedom in the reionization history,
and be compatible with current observations, 
we use a model characterized by one additional parameter (hereafter addressed as Poly-Reion) defined as:
\begin{equation}
 x_e(z)=(1+F_{\rm He})f(z),
\end{equation}
where $f(z)$ is a polynomial. $f(z)$ is a Piecewise Cubic Hermite Interpolating Polynomial (PCHIP). 
The polynomial is determined by the electron fraction at four redshifts (the nodes in the polynomial): $z=0,5.5,7$ and $z_{xe=0}$. 
From the observations we have $f(z=0)=f(z=5.5)=1$ ~\cite{Fan:2006dp} conservatively~\footnote{The authors report neutral hydrogen fraction to be $10^{-4}$ at $z<5.5$. At the same time since they find a sharp increase in 
neutral hydrogen up to $0.1$ by redshift 6, using this Poly-Reion method we will be allowing such scenarios with some neutral hydrogen before $z=5.5$.}, thereby we are excluding 
too much late reionizations. 
$f(z=7)=x_e(z=7)$ is a free parameter,where the choice of a node at $z=7$ is well justified by the observed presence of neutral hydrogen 
at different lines of sight using Quasar and Gamma Ray Burst data~\cite{Bouwens:2015vha}. 
$z_{xe=0}$ defines the epoch in the past where the Universe was fully neutral and it is fixed
by solving for the given optical depth: 
\begin{equation}
 \tau=\int \sigma_{\rm T} n_e(z) dl,~\label{eq:reion-los}
\end{equation}
where, $n_e(z)$ is the free electron density and $\sigma_{\rm T}$ is the Thomson scattering cross section.
The $f(z)$ can never be greater than 1 or less than 0 and respects a monotonic increase in free electron density from past to present. Using the 
$f_{z}$'s at the given nodes, PCHIP ensures the monotonicity of $x_e(z)$. 
The Poly-Reion model has three main advantages with respect to the step model:
\begin{itemize}
\item it is by construction in agreement with observations of a fully ionized medium at $z=5.5$;
\item it allows more freedom being characterized by the free parameter $f(z=7)=x_e(z=7)$;
\item it naturally provides the redshift where reionization started as a derived parameter.
\end{itemize} 

Unlike the Tanh model, the Poly-Reion does not lead to symmetric reionization histories. 
Although there may be significant overlaps in the reionization histories they model, 
the Tanh-Reion parametrization is not nested within the Poly-Reion model. The Poly-Reion is also different from the 
asymmetric reionization history analyzed with Planck data in~\cite{Adam:2016hgk} 
and in the perspective of the CORE concept in~\cite{core:cosmoparam}.
Note that in a recent paper, two of us have discussed a Poly-Reion form of reionization history~\cite{Hazra:2017gtx}. 
The model discussed here is nested within that generic model. 

\subsection{Solving for reionization}

The Poly-Reion reionization history has an extra parameter $x_e(z=7)$ in addition to $\tau$, which is the only free parameter 
for Tanh-Reion.
The earliest redshift that is used for the polynomial is $z_{xe=0}$ and we require $f(z)=0$ 
(for all $z\ge z_{xe=0}$) such that the Universe is fully neutral before that redshift. 
We do a search from $z=7.25$ (we use this value instead of $7$ 
to avoid sharp step for CAMB integrals) and the maximum redshift 
for the onset of reionization (we assume $z_{\rm max}=70$) that matches the $\tau$ provided. 
$x_e(z=7)$ is varied between 0 to 1, the values in this range 
which cannot solve for the values of $\tau$ are rejected. 

First we derive the constraints for the Planck 2015 data. 
We use the public available low and high-$\ell$ likelihood Planck 2015 TTTEEE + lowTEB + lensing.
The Planck lensing likelihood lowers the value of the optical depth, using Tanh-reion, 
to $\tau = 0.063 \pm 0.014$, 
which is closer to the most recent value reported in Planck 2016 Intermediate paper with $\tau = 0.0596 \pm 0.0089$~\cite{Aghanim:2016yuo}.
We use the mean values as fiducial parameters for the CORE simulated data.

%%%%%%%%%%%%%%%%%%%%%%%%%%%%%%%%%%%%%%%%%%%%%%%%%%%%%%%%%%%%%%%%%%%%%%%%%%%%%%%%%%%%%%%%%%%%%%%%%%%%%%%%%%%%%%%%%%%%%

\subsection{Constraints on Reionization history}

As already been mentioned, the Tanh-Reion is not nested within the more general Poly-Reion.
However, since we aim to propose Poly-Reion as an alternative model 
for reionization history, 
we will present and discuss the results 
of the constraints on cosmological parameters using the two models.

\renewcommand{\arraystretch}{1.1}
\begin{table*}[!htb]
\begin{center}
\vspace{4pt}
\begin{tabular}{| c | c | c |}
\hline\hline
\multicolumn{3}{|c|}{\bf Parameter Constraints for Poly-Reion}\\
\hline
Parameters & Planck& CORE\\
\hline
 $\Omega_{\rm b}h^2\times10^{2}$ 
 &$2.23\pm0.03$
 &$2.228\pm0.007$\\
\hline
 $\Omega_{\rm CDM}h^2\times10^{2}$ 
 &$11.89^{+0.26}_{-0.27}$
 & $11.885^{+0.054}_{-0.055}$\\
 \hline
 $\tau$ 
 &$0.069\pm0.025$
 & $0.0696\pm0.0044$  \\
 \hline
 $x_e(z=7)$ 
 &$0.45^{+0.49}_{-0.37}$
 &$0.53^{+0.33}_{-0.29}$\\
\hline
 $H_0$ 
 &$67.65^{+1.23}_{-1.18}$
 & $67.68^{+0.23}_{-0.22}$ \\
 \hline
\hline
$\ln[10^{10} A_{\rm S}]$ 
&$3.07^{+0.047}_{-0.046}$
& $3.071^{+0.0076}_{-0.0074}$ \\
\hline
$n_{\rm S}$ 
&$0.966\pm0.009$
&$0.9661\pm0.0029$ \\
\hline
\hline
 $\Delta_{z}^{\rm Reion}$ 
 &$8.2^{+7.2}_{-7.3}$
 &$8.5^{+2.2}_{-3.1}$  \\
\hline
 $z_{\rm re}$ 
 & $7.25^{+2.4}_{-1.3}$ 
 & $7.46^{+1.86}_{-1.34}$  \\
\hline
 $\sigma_8$ 
 & $0.818^{+0.017}_{-0.016}$ 
 & $0.819\pm0.002$\\
\hline\hline
\end{tabular}
\end{center}
\caption{~\label{tab:bounds-reion}Present and projected constraints on reionization history (and other background and perturbation parameters) assuming Poly-Reion, using Planck data and CORE 
simulated data respectively. Errors correspond to 95\% C. L. A comparison with Planck results in~\cite{Planck:2015Param} indicates that apart from the
optical depth, all other parameters have similar constraints.} 
\end{table*}

The best fit $-2\ln{\cal L}$ obtained from Powell's BOBYQA method~\footnote{In order to obtain the best fits we use Powell's BOBYQA 
(Bound Optimization BY Quadratic Approximation) method of iterative minimization~\cite{powell}.}
for Planck TTTEEE + lowTEB + lensing when Tanh-Reion is used is 12947.2. For Poly-Reion we find $-2\ln{\cal L_{\rm max}}=12944.6$. 
Hence we find roughly an improvement of 2.6, with an extra parameter, compared to the Tanh reionization histories. 
%This improvement may mainly be due to ther more flexibility at the power spectrum level allowed by the Poly-Reion 
%model in contrast with the more rigid Tanh-Reion model.  
%This one extra parameter is allowing
%an extended reionization model and thereby allowing a bit more flexibility to the CMB angular power spectrum resulting in a slightly better fit. Hence it is evident that 
%our model of reionization is appropriate to use as an alternative. 
In table~\ref{tab:bounds-reion} we provide the bounds on the cosmological parameters when
Poly-Reion is used for the reionization history for both Planck real data and CORE forecasts. 
Instead of presenting the results for $z_{xe=0}$ we prefer to use 
the duration of reionization $\Delta_{z}^{\rm Reion}$ which is defined 
as the difference between the redshift when the reionization is 10\% complete and the redshift when it is 99\% complete.
This parameter is in fact much better constrained by both Planck 
and CORE with a much better behaved distribution compared to $z_{xe=0}$.
Note that this duration of reionization is different from the $\delta_{\rm Reion}$ used in Eq.~\ref{eq:reion-tanh}.\\
%We find since the parametrization allows reionization histories
%with $x_e(z)$ extending to higher redshifts with very long tails with a very small value, compared to $z_{xe=0}$ the datasets (Planck and of course CORE) can constrain $\Delta_{z}^{\rm Reion}$
%significantly better. Hence we provide $\Delta_{z}^{\rm Reion}$ constraints instead.

\begin{figure*}[!htb]
\begin{center} 
\resizebox{160pt}{115pt}{\includegraphics{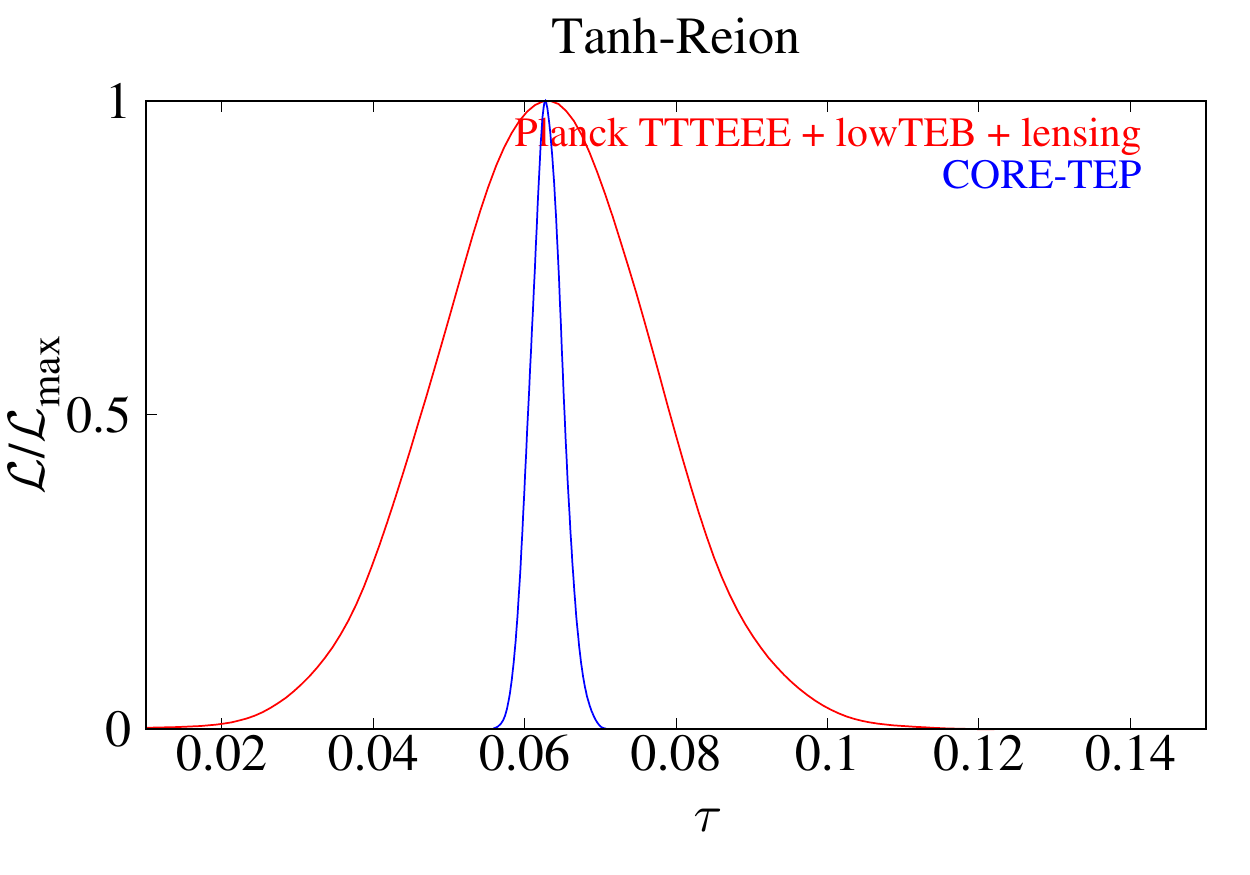}} 
\resizebox{160pt}{115pt}{\includegraphics{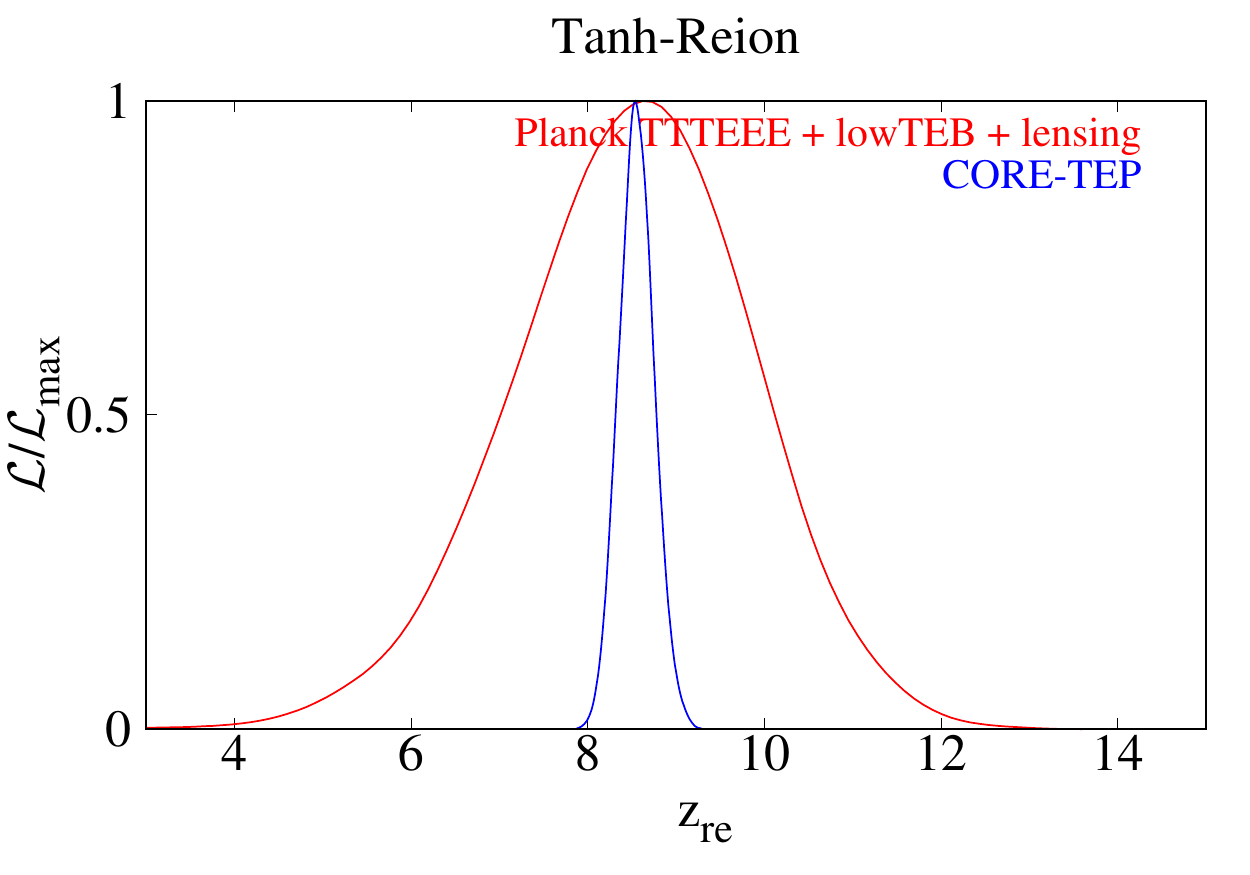}} 

\resizebox{160pt}{115pt}{\includegraphics{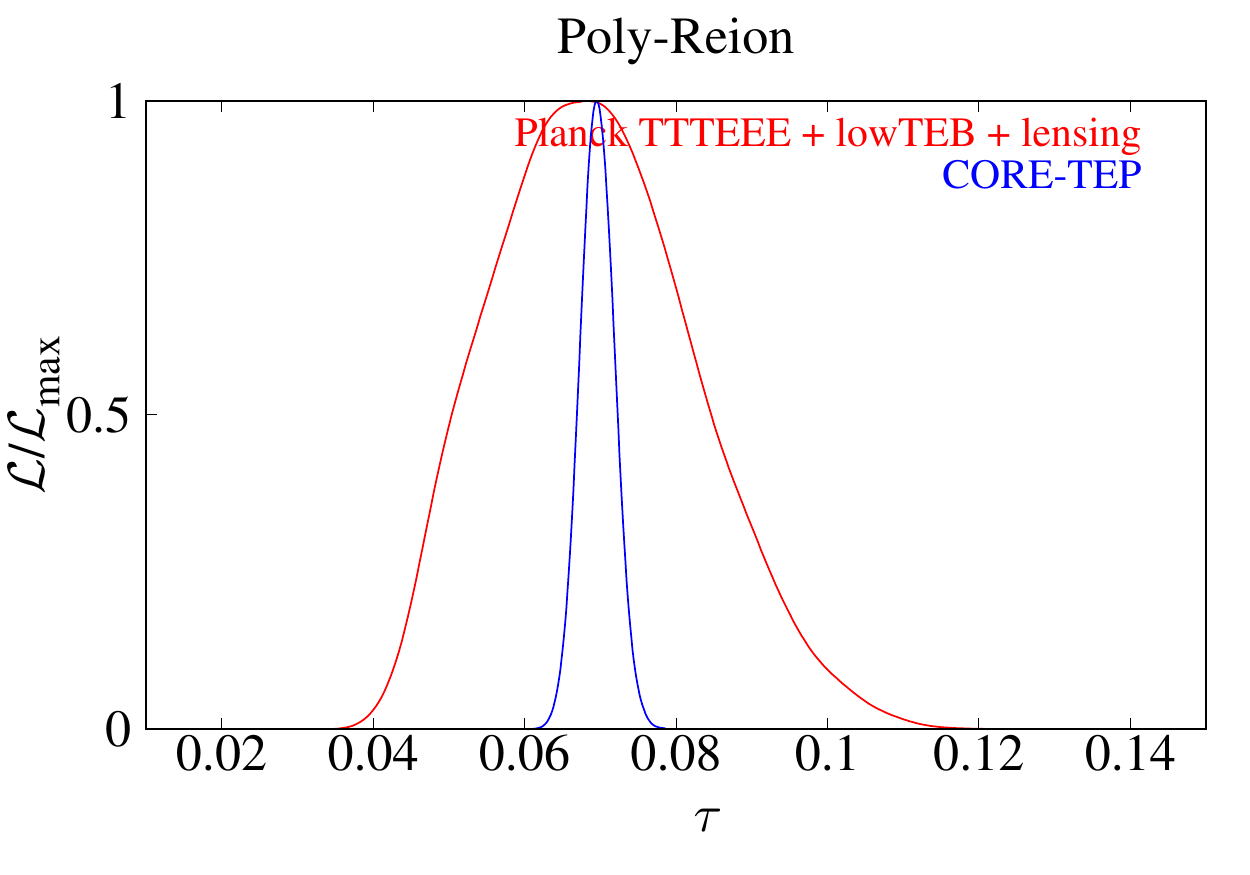}} 
\resizebox{160pt}{115pt}{\includegraphics{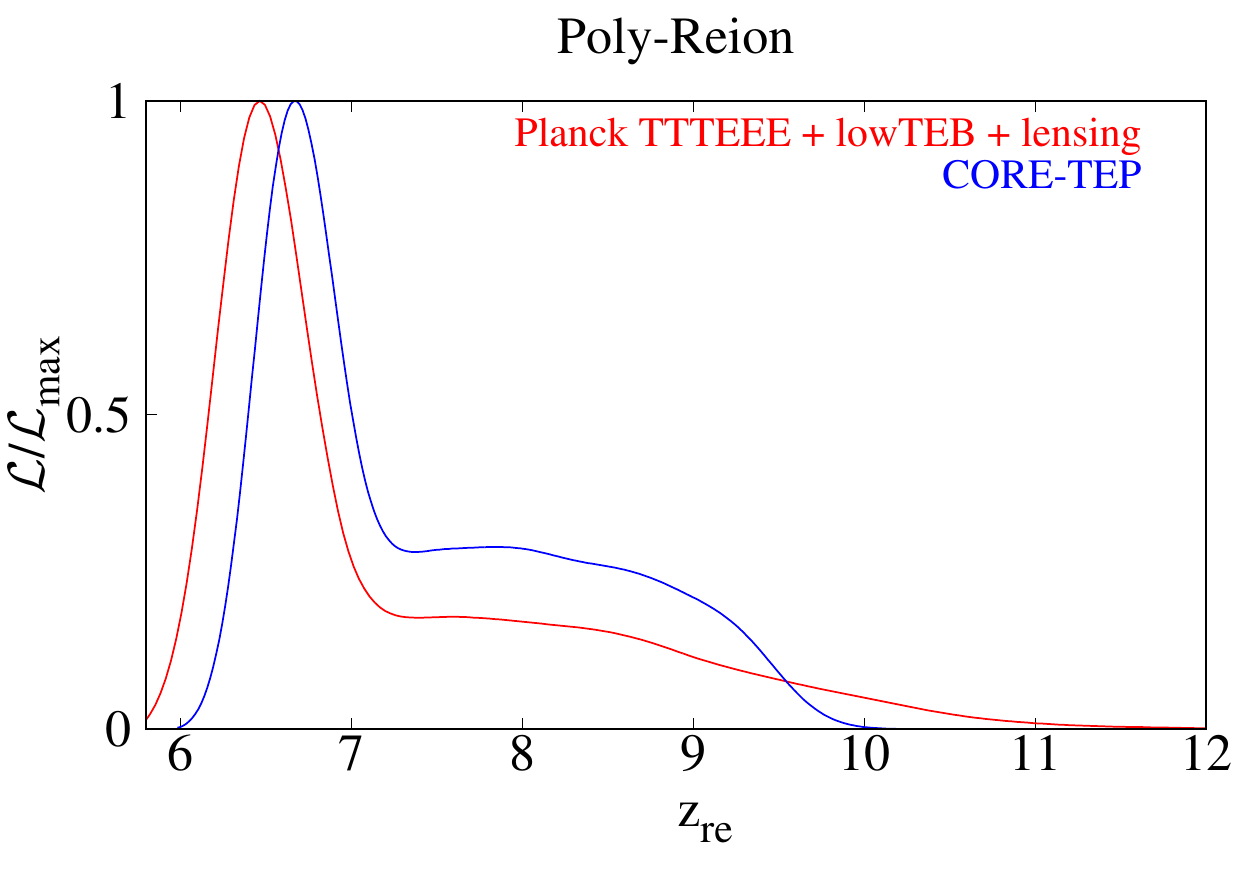}} 

\resizebox{160pt}{115pt}{\includegraphics{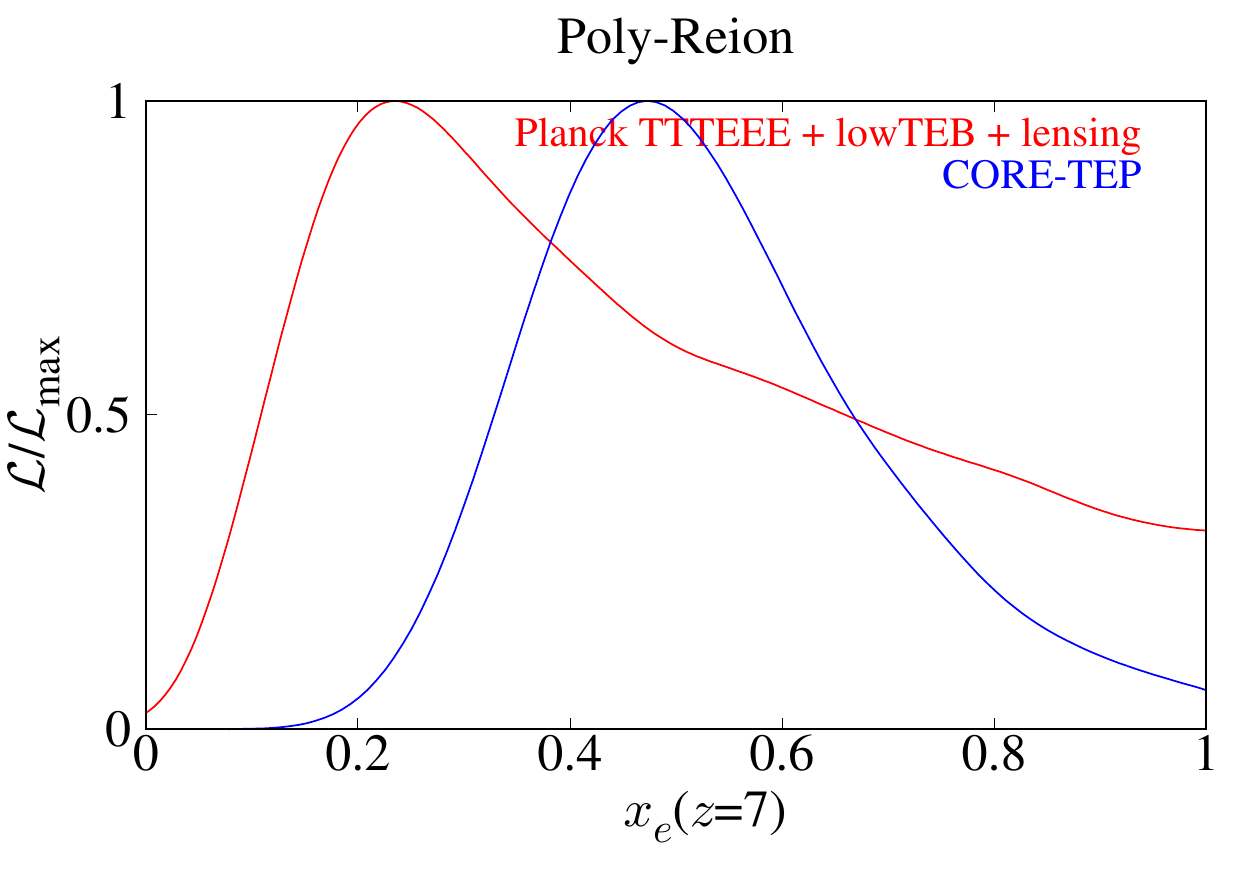}} 
\resizebox{160pt}{115pt}{\includegraphics{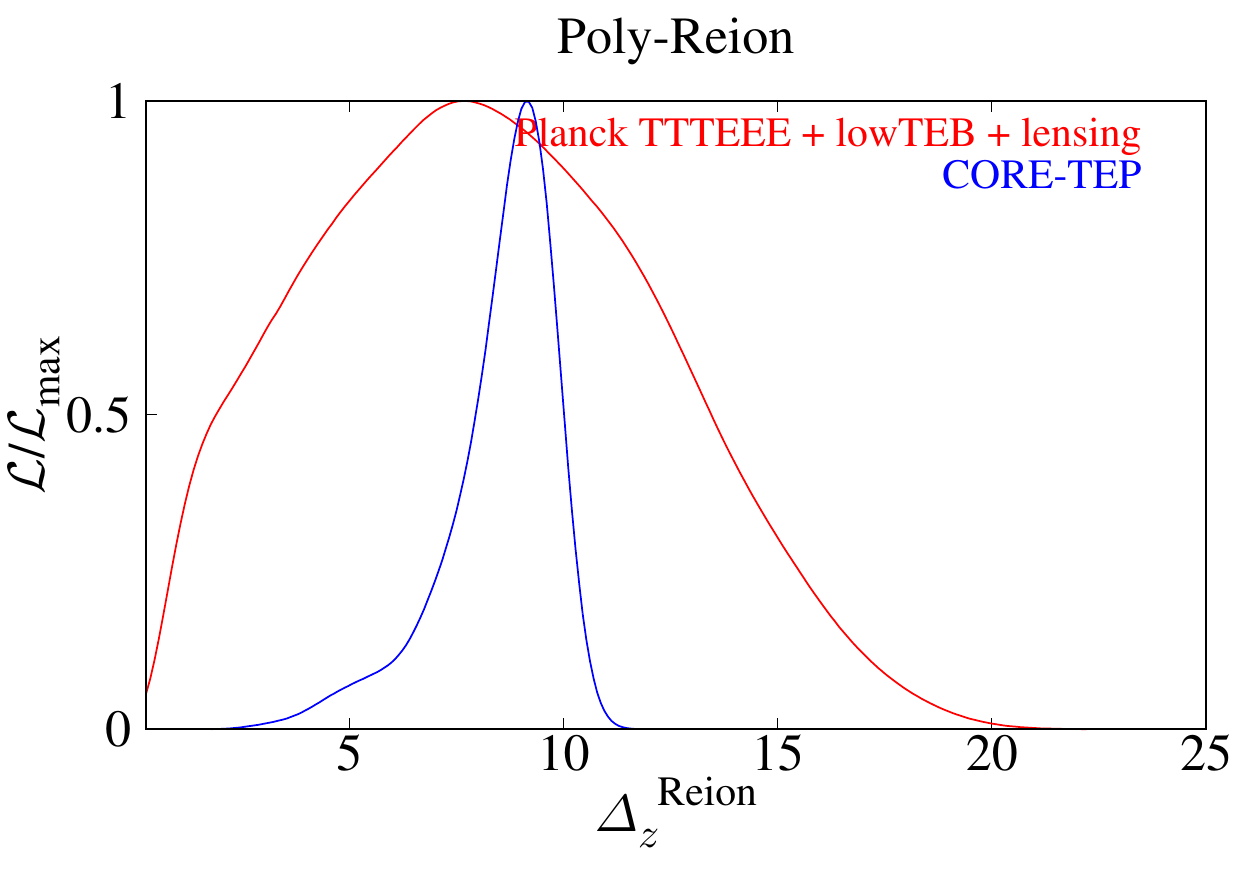}} 

\resizebox{180pt}{140pt}{\includegraphics{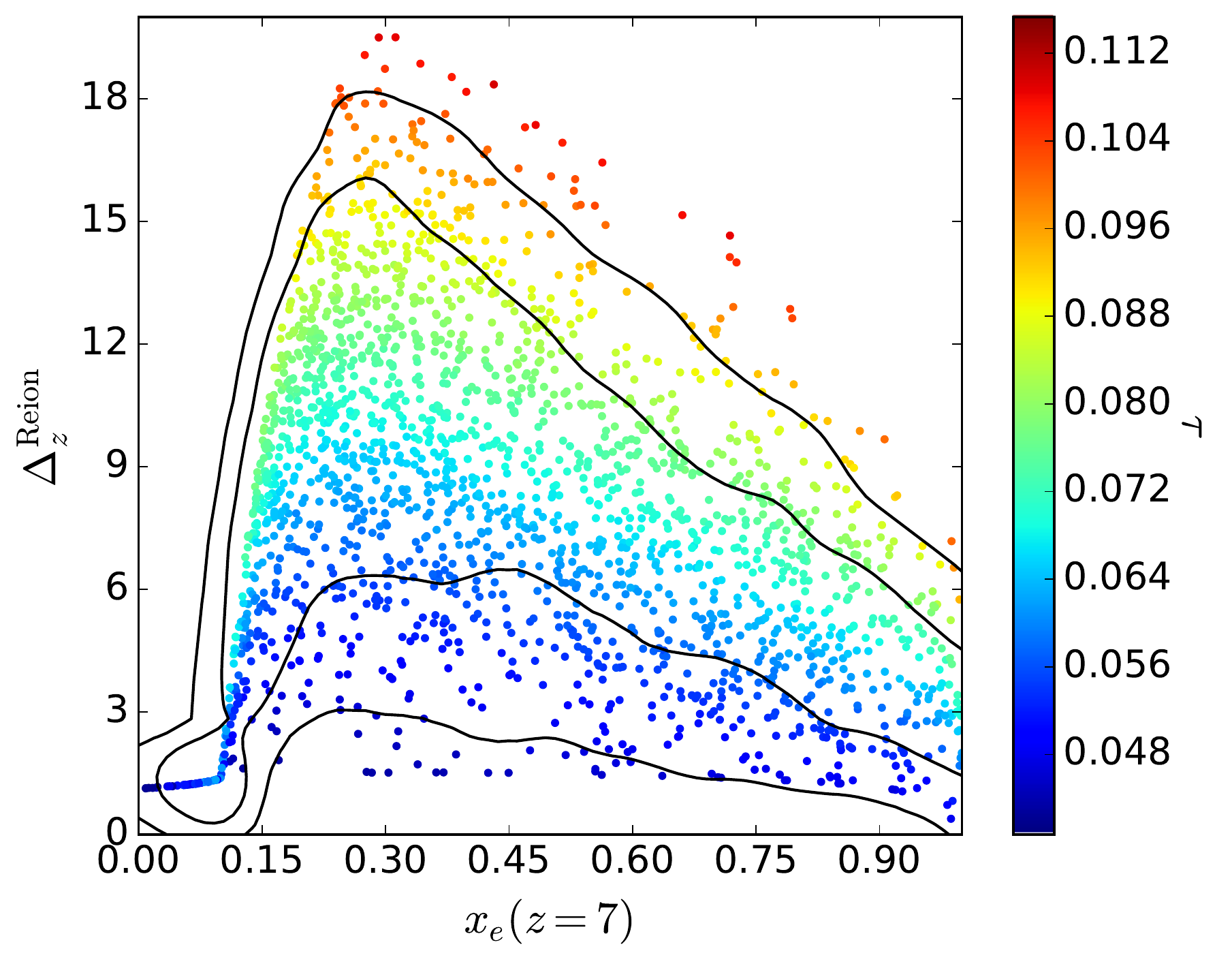}} 
\hskip -8pt\resizebox{180pt}{140pt}{\includegraphics{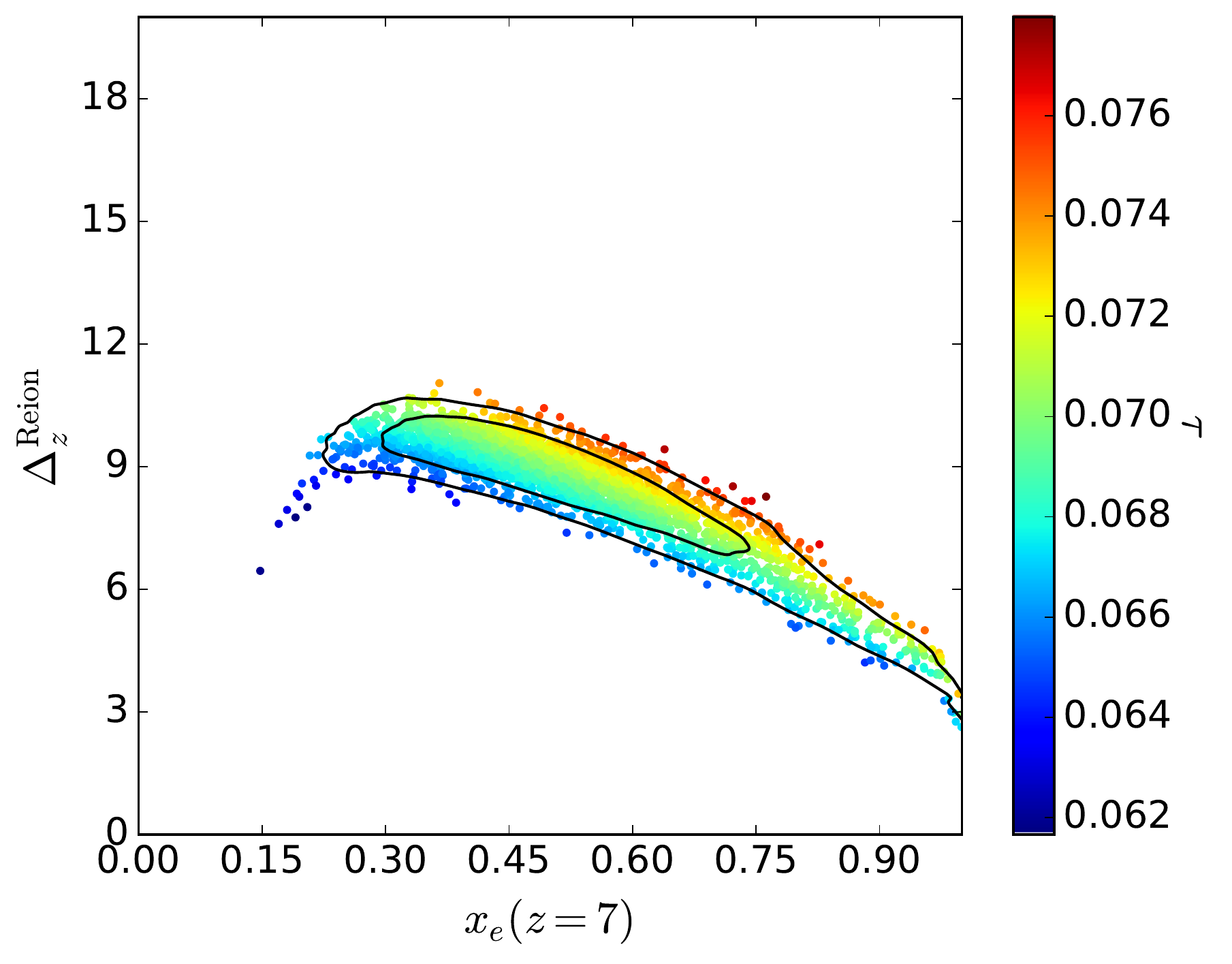}} 
\end{center}
\caption{\footnotesize\label{fig:params_reionhist}Current constraints on the reionization history and the forecasts for CORE. First three rows from the top
represent the constraints from Planck TTTEEE+lowTEB+lensing in red and the projected CORE constraints in blue. In the topmost panel we have plotted the optical depth ($\tau$) and 
the reionization redshift ($z_{\rm re}$) for the Tanh model of reionization histories. The rest of the rows represent Poly-Reion scenario. Two middle rows plot the marginalized
probability distribution of $\tau,z_{\rm re},x_e(z=7)$ and reionization duration $\Delta_{z}^{\rm Reion}$ respectively (in order of reading from left to right and top to bottom).
The plots at the bottom most panel show the correlation between history parameters $x_e(z=7)$ and $\Delta_{z}^{\rm Reion}$. The left plot shows the Planck bounds and the right
plot predicts the CORE constraints.
Note that, as expected, CORE predictions give significantly tighter constraints on the history of reionization.}
\end{figure*}

We note by comparing the results in table~\ref{tab:bounds-reion} 
and the Planck results~\cite{Planck:2015Param} that the constraints 
on background and power law PPS parameters do not change much for Tanh-Reion and Poly-Reion, instead the mean 
values for $\tau$ and $A_{\rm S}$ are shifted due to degeneracies. 
We present also the two derived parameters: $z_{\rm re}$ which represents 
the redshift where the electron fraction from hydrogen is exactly half and $\Delta_{z}^{\rm Reion}$ which is 
duration of reionization.
In figure~\ref{fig:params_reionhist} we present the comparison of the reionization histories from Planck and the forecasts
for CORE. In order to display the correlation, we provide the 2D marginalized contours for $x_e(z=7)$ and $\Delta_{z}^{\rm Reion}$
at the bottom panel of the same figure. The dependence on $\tau$ is represented by the rainbow colormap of the points.

\begin{figure*}[!htb]
\begin{center} 
\resizebox{210pt}{140pt}{\includegraphics{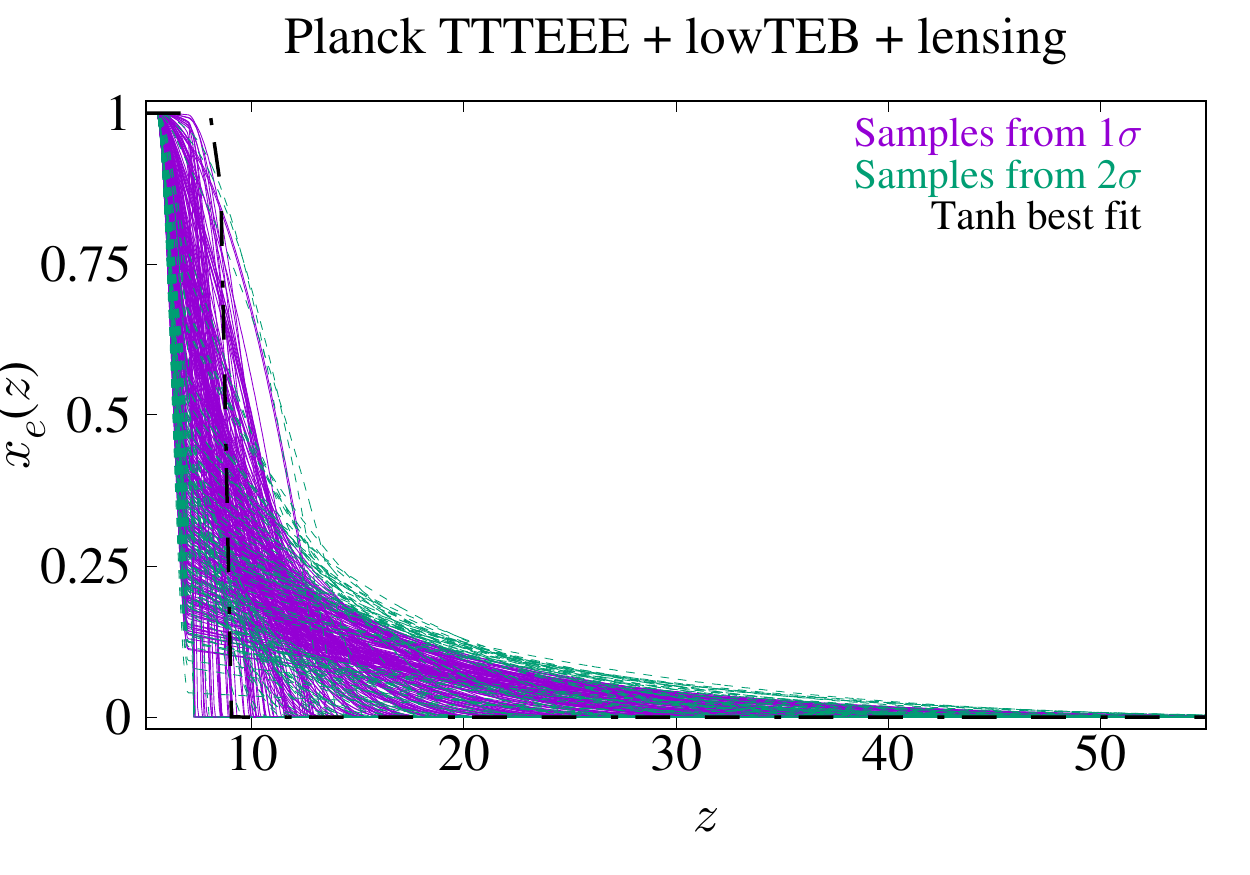}} 
\resizebox{210pt}{140pt}{\includegraphics{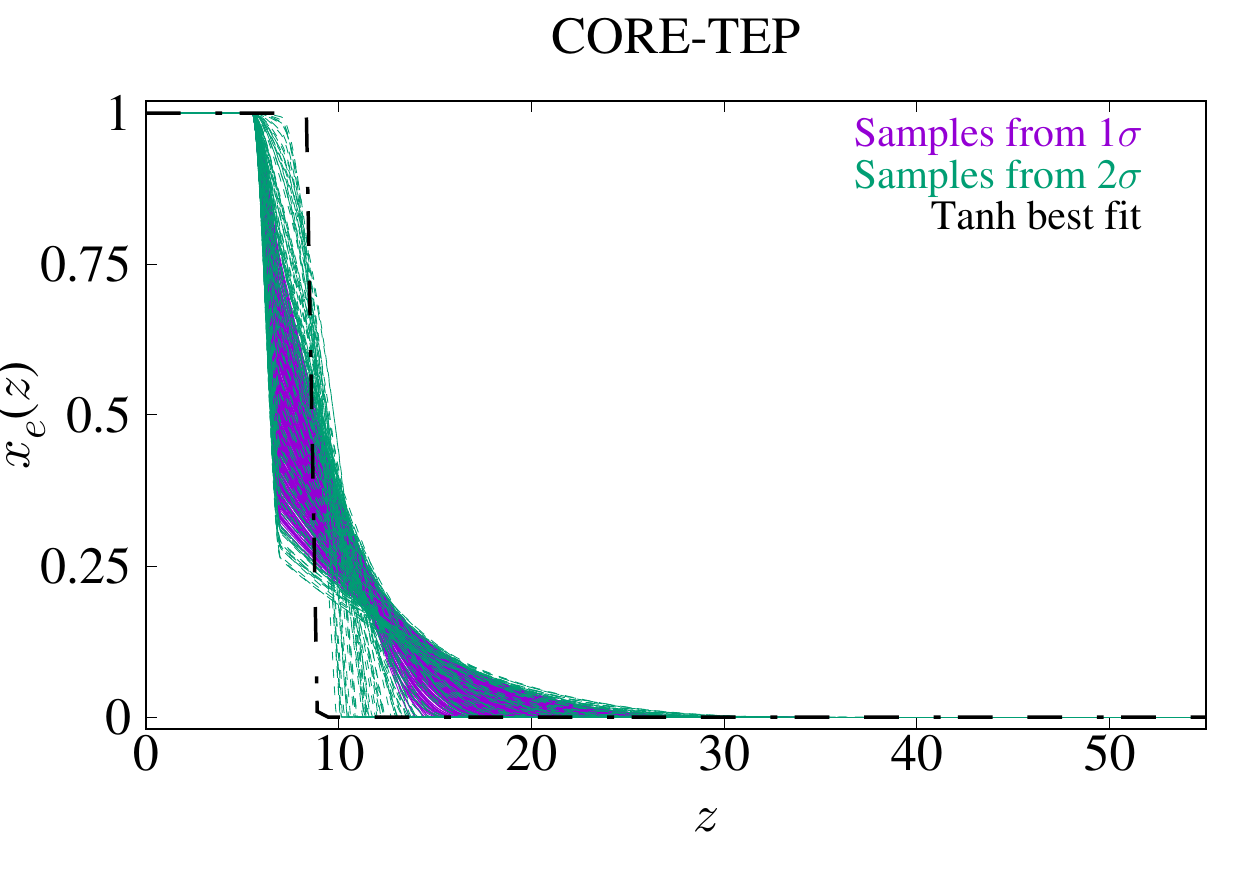}} 
\end{center}
\caption{\footnotesize\label{fig:reionhist}Free electron fraction (not including electrons from singly and doubly ionized Helium, while in the analysis we have included both) as a function of redshift.
The samples are plotted for the actual Planck (left) dataset and the simulated CORE (right) dataset. Samples from 1$\sigma$ confidence are plotted in 
solid and 2$\sigma$ samples are plotted in dashed lines. Needless to mention, these lines integrate to provide different values of the optical depth. 
We have plotted samples instead of bands on the electron fraction as some of the features and the flexibility of the model are better seen in sample representation. In black dash-dotted line, we have plotted 
the best fit Tanh-Reionization history obtained with Planck TTTEEE + lowTEB + lensing.}
\end{figure*}

For a given electron fraction, to solve Eq.~\ref{eq:reion-los} for 
increasing value of optical depth, we need to increase the total 
duration of reionization period and thereby we require higher (earlier) value for $z_{xe=0}$. 
To show the different reionization histories given by this model we 
have randomly selected from the samples within 1$\sigma$ 
and between $1-2\sigma$ contours 
and we plot the electron fraction (not considering Helium reionization) 
for the entire reionization histories for Planck and CORE bounds in figure~\ref{fig:reionhist}
in the redshift range $0-55$. For comparison the best fit of Tanh-Reion 
is plotted in dashed black. Since samples from different confidence limits 
will intersect each other in redshifts, we plot the samples rather than providing confidence bands.
Within the chosen fiducial models we note how Planck 
constraint is limited to the range $5.5<z<50$ (the lower limit being prior dominated) CORE can reduce the range to $5.5<z<25$.
Though this constraint is strictly dependent on the chosen fiducial reionization history, the improvement due the 
better noise sensitivity in CORE is quite evident. By comparing table~\ref{tab:bounds-reion} and Eq.~\ref{Eq:PlanckCoreArray}, we see that 
$\tau$ uncertainty does not degrade significantly at CORE sensitivity by allowing more freedom in the reionization history, 
as was already notice in~\cite{core:cosmoparam} for a different asymmetric reionization history. 

We note how although Tanh-Reion and Poly-Reion both provide very similar fit to the data, 
they have very different reionization histories, allowing also for mutually exclusive results. 
In both plots, the Poly-Reion 1-2$\sigma$ samples extend to the direction of lower electron 
densities at low-$z$ and thereby demanding a higher 
value for the beginning redshift of reionization. 
This signifies that there are hints and possibilities of extended reionization. A free-form reionization history parametrization 
as in~\cite{Hazra:2017gtx} will be more effective in future in understanding the redshift dependent constraints of the free electron 
fractions and that will allow complicated models relaxing these bounds to certain extent.

\section{Confusion between inflationary features and extended reionization history}~\label{sec:degen}

We now discuss to what extent an extended reionization scenario such as Poly-reion can 
be partially degenerate in generating similar polarization anisotropy as the large scale inflationary
features in the PPS produce. 
In order to asses this degeneracy, we use the following approach. 
WWI-A, amongst all the best fits, generates a suppression and large scale oscillation and thereby provides 
the improvement in fit to the TT data. Though the large scale polarization data at hand can not support or
deny this feature, we assume WWI-A to be the fiducial model of the Universe. 
We present directly the forecasts for CORE, that with its good polarization sensitivity may target this issue.
Being characterized by one extra parameter, while Poly-Reion is not expected to
provide sufficient flexibility to mimic the polarization features from WWI-A, it can produce a variety of extended reionization 
histories. For completeness we use the TT, TE 
and EE data together even if Poly-Reion is expected to affect mainly the large scale polarization. 

To investigate the possible confusion caused by this degeneracy, we study how the fit to the data where the fiducial is given by the WWI-A is improved when using a power 
law PPS (hence not the real PPS model of the Universe we are assuming)
with Poly-Reion instead of Tanh-Reion. In this way we may asses if the use of Poly-Reion model may lead to a better fit with a power 
law PPS though the real Universe is represented by a WWI-A cosmology reducing the possibility to detect such a model in more free reionization histories.

\begin{figure*}[!htb]
\begin{center} 
\resizebox{280pt}{200pt}{\includegraphics{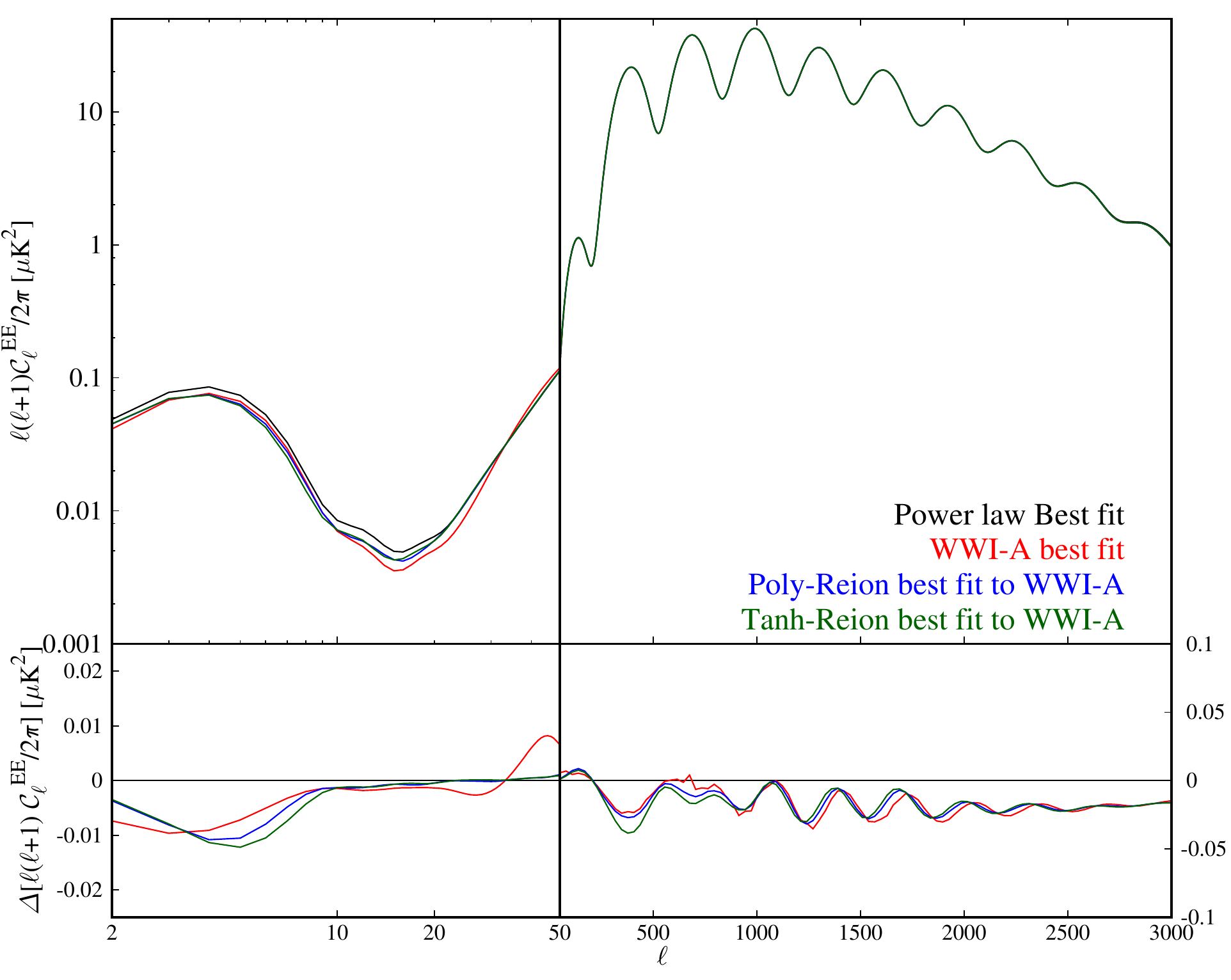}} 
% \hskip -8pt\resizebox{220pt}{200pt}{\includegraphics{plots/clTE_Reion_WWIA_TE-eps-converted-to.pdf}} 
\end{center}
\caption{\footnotesize\label{fig:degen_cldiff} Polarization angular power spectra for different models. The angular power spectrum 
are plotted for the best fit Planck results for power law and WWI-A. For the Tanh-reion and Poly-reion, the plotted power spectra 
represent the best fit to the simulated WWI-A spectrum for CORE. The bottom inset plot represent the differences in the angular 
power spectra {\it w.r.t} the power law best fit. Note that though the Poly-reion does not have much flexibilities that can address
the inflationary features in the polarization data, in both TE and EE spectra, it performs better than the Tanh-reion, owing to its 
one extra parameter.}
\end{figure*}
We compare the difference in average likelihoods between the power law model 
with Tanh-Reion and Poly-Reion compared to WWI-A model with Tanh-Reion. We simulate CORE spectra
with best fit WWI-A feature and Tanh-Reion.
We derive the $\chi^2=-2\ln{\cal L}$ for (i) the WWI-A with Tanh-Reion, (ii) power law PPS with Tanh-Reion 
and (iii) power law PPS with Poly-Reion all to fit WWI-A simulated spectra. We then compare the average 
 $\chi^2$ within 1-10\% of the maximum likelihood values. 
We find, if we use Poly-Reion model with power law, it can fit the WWI-A feature 6-9\% better (in $\chi^2$) compared to Tanh-Reion
case. Therefore we can confirm the Poly-Reion introduces features in reionization history that can mimic parts of the WWI-A feature 
to a limited extent. 

In the figure~\ref{fig:degen_cldiff}, we plot the differences between (i),(ii) and (iii) in the EE anisotropy power spectra.
The best fit spectra for power law and Tanh-Reion and from (i),(ii) and (iii) are plotted in black, red, green and blue respectively. In the inset, we plot the differences between 
different runs and the power law best fit spectra. Note that in both EE and TE spectra,(iii) is closer to (i) compared to (ii) at large scales. At smaller scales the changes are 
due to difference in the best fit background cosmological parameters.

\begin{figure*}[!htb]
\begin{center} 
\resizebox{220pt}{160pt}{\includegraphics{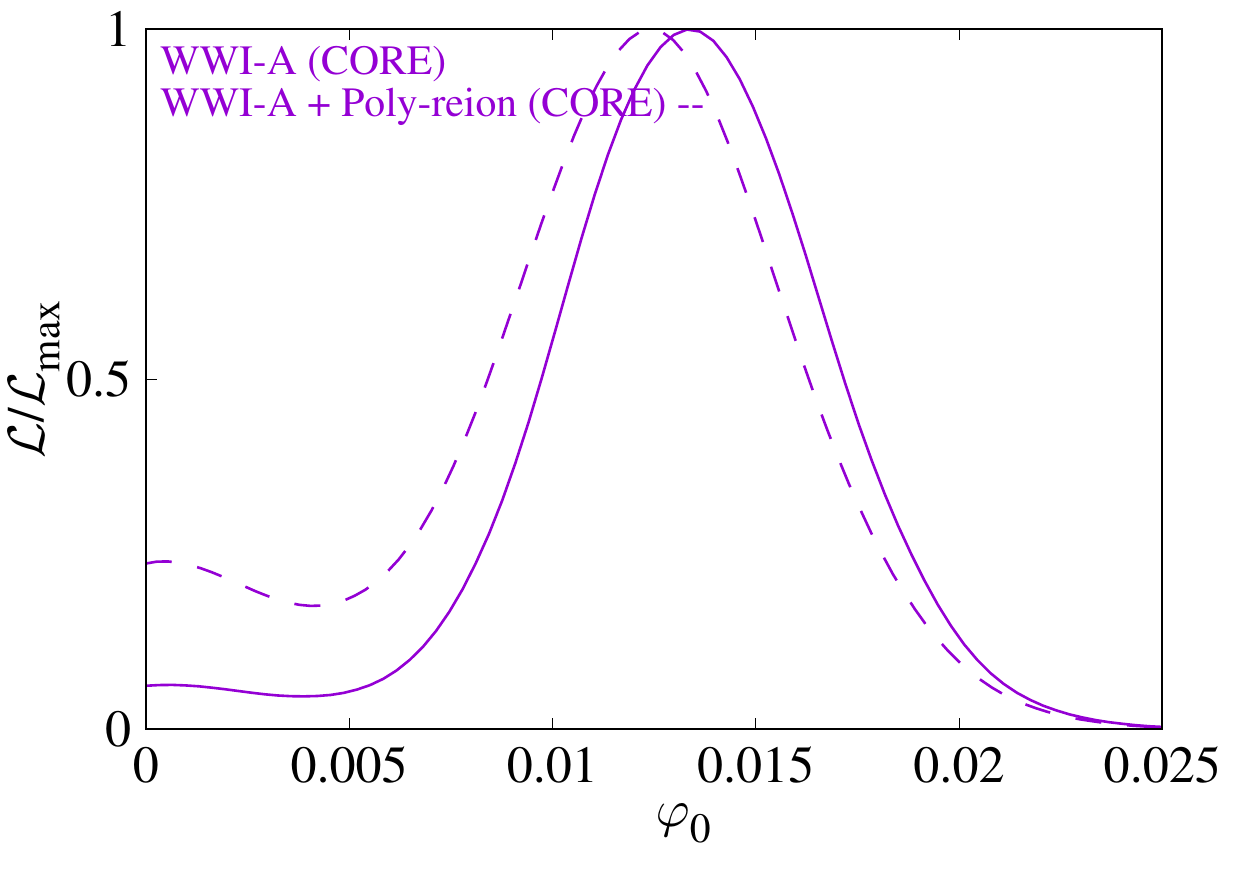}} 
\hskip -8pt\resizebox{220pt}{160pt}{\includegraphics{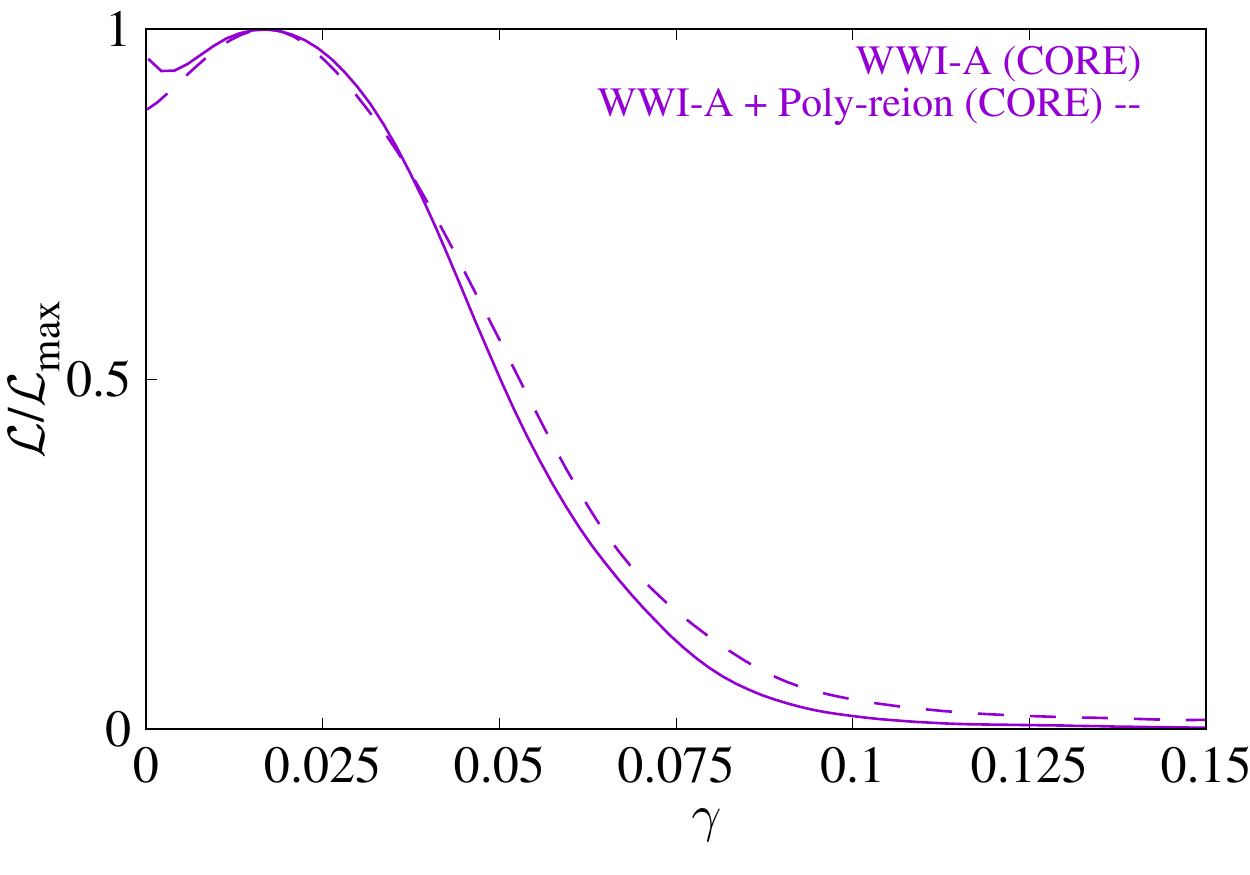}} 

\resizebox{220pt}{160pt}{\includegraphics{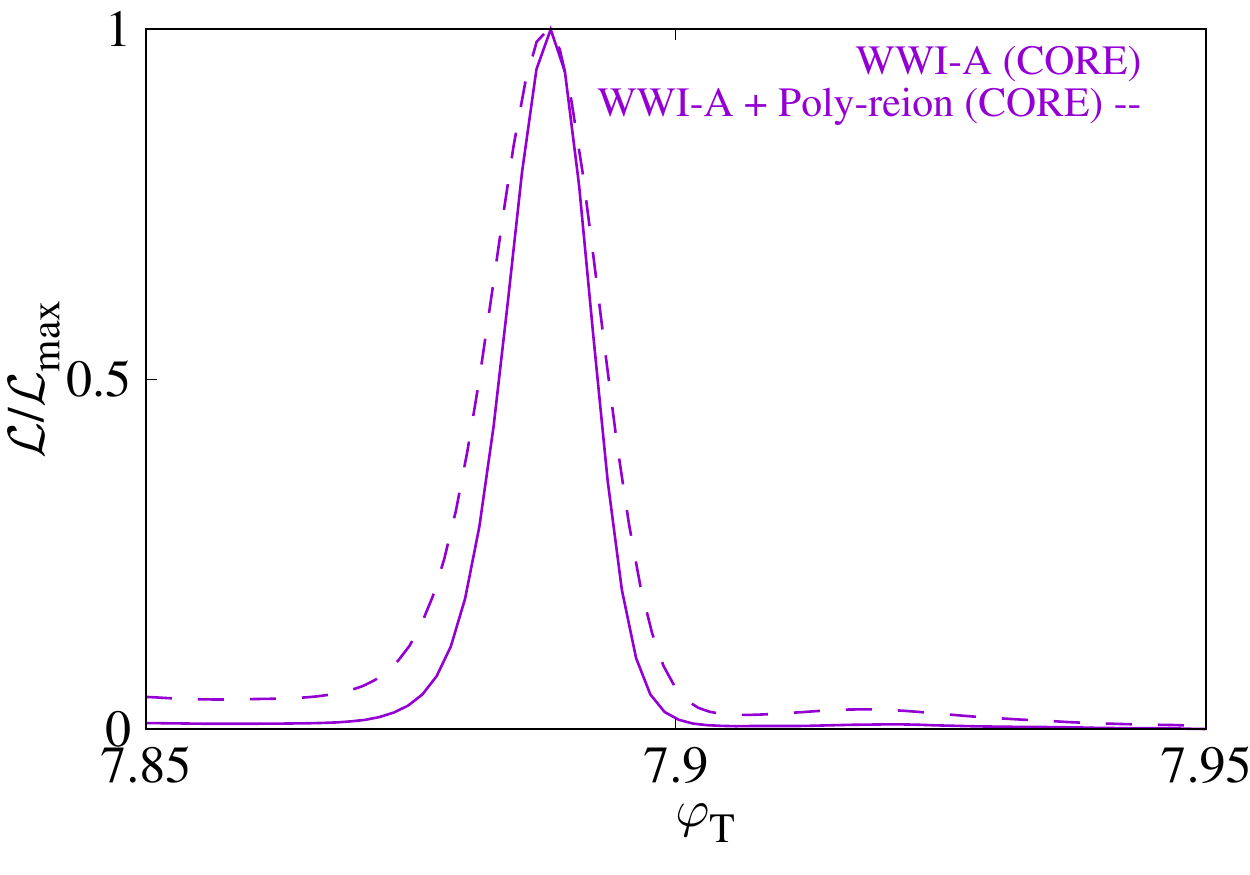}} 
\hskip -8pt\resizebox{220pt}{160pt}{\includegraphics{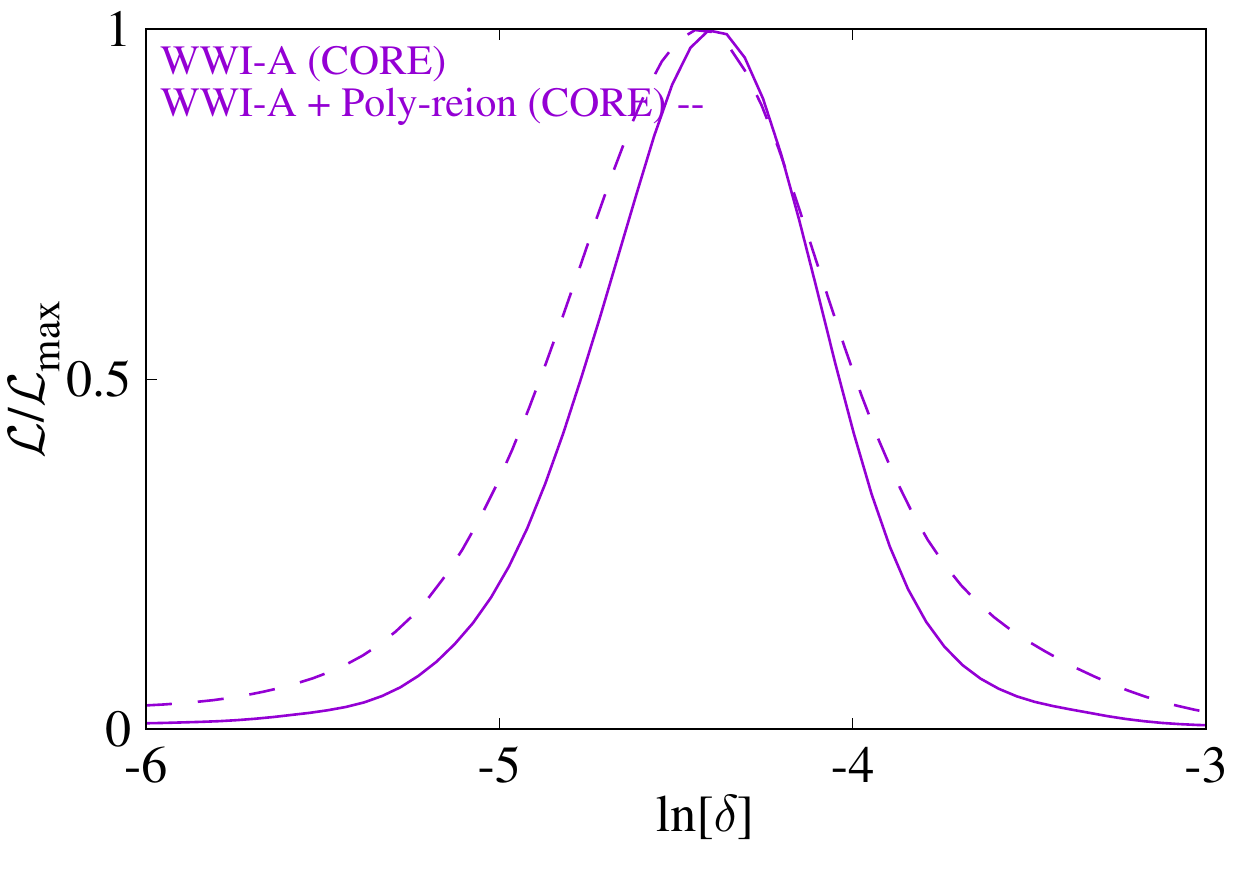}} 
\end{center}
\caption{\footnotesize\label{fig:inf-degen}Shifts in the inflationary potential parameters when Poly-reion is used as model for reionization instead of Tanh-reion. We have used WWI-A
as fiducial model as this best fit has features only at largest scales. We note that the parameter $\phi_{0}$ which is moderately constrained ($2-3\sigma$) in the Tanh-reion scenario, is now 
constrained with slightly less significance ($1-2\sigma$). Negligible change in the parameter responsible for large scale suppression ($\gamma$) is noticed. Since the position ($\phi_{\rm T}$) and the width ($\delta$) of the 
discontinuity is related with the extent of discontinuity ($\phi_{0}$), we notice change in these parameters as well. The changes in the background parameter constraints are negligible and 
therefore not presented.}
\end{figure*}

To investigate further the extent of the confusion, we attempt to fit WWI-A fiducial with WWI model combined with Poly-reion scenario. Here we use temperature, polarization and lensing 
likelihood similar to what we have considered in section~\ref{sec:wwiforecast}. We compare the obtained marginalized likelihoods of the inflationary feature parameters ($\phi_{0},\gamma,\phi_{\rm T}$ and 
$\ln[\delta]$ with the likelihoods from section~\ref{sec:wwiforecast}) in figure~\ref{fig:inf-degen}. We compare the cases where WWI-A is used as fiducial. Note that we do not find significant
change in the posteriors. Definitely the significance of the feature has been reduced as for
$\phi_{0}$ posterior has shifted closer to {\it zero}. We find that the previously obtained 
moderate significance $2\sigma-3\sigma$ reduces to $1\sigma-2\sigma$ when we use Poly-reion. This decrease in significance is partially caused by the extended reionization 
that can generate parts of the WWI-A feature in the polarization anisotropy. We find no noticeable change in the background parameters.

This study of confusion here indicates that an extended reionization history will not restrict our capabilities of finding large scale features to significant extent with 
the future surveys having the sensitivity of CORE. We have used more nodes in the Poly-reion parametrization between $7<z<70$ to allow more complex reionization 
histories, however, we do not find notable increase in degeneracy between the reionization history and the WWI-A. Therefore we can further comment that 
given a realistic reionization history with $0<x_e(z)<1$ and monotonic increase of $x_e$ with the decrease in redshifts, the degeneracy between large scale inflationary
features and the process of reionization is limited.

\section{Conclusions}~\label{sec:conclusions}

In this paper, we address three important aspects of cosmological parameter estimation with next generation space based 
CMB surveys, namely, features in the primordial power spectra,
non-instantaneous reionization histories and the possible degeneracies between these two aspects. 
For the first part, to avoid the use of different models of features, we use the more general Wiggly 
Whipped Inflation that provides different feature models within the same framework. 
WWI includes inflationary phase transitions with discontinuities in potential or in its derivative. 
In our earlier analysis~\cite{HSSS:2016}, we had demonstrated that using only one potential 
with 2-4 extra parameters, we can generate multitudes of inflationary features that are discussed 
in literature using different potentials. 
The advantage of using WWI is that with the same potential or same framework we can locate the 
existence of features that are supported by the data in one step.

Using Planck 2015 temperature and polarization data, from WWI framework we identified 5 distinct features 
that improve the fit to the Planck data, although not at a statistical significance level, with 
respect to the power law form of the PPS: WWI-[A, B, C, D] and WWI$'$. 
WWI-[A, B, C, D] come from the same potential with a discontinuity whereas WWI$'$ is from another 
potential with discontinuity in the derivative. 
All the PPS considered here contain suppression 
on large angular scales then we have that: WWI-A contains localized wiggles 
in the PPS at scales corresponding to $\ell=20-40$; WWI-B, C and WWI$'$
introduce oscillations which are extended towards smaller scales compared to WWI-A; 
WWI-D contains non-local oscillations that continue up to smallest angular 
scales (up to the observational limit assumed in this paper).

In this paper we addressed the possibility that future space CMB observations
will detect features in the inflaton potential if they represent the true model of the Early Universe. 
We take CORE as an example for a future CMB space satellite concept 
and we tested WWI framework against simulated data for temperature, 
polarization and lensing potential. 
The results show that broadly we have a moderate significance
to features in the inflaton potential leading to features in the primordial power spectrum
that are localized to certain cosmological scales.
While for WWI-A and WWI-B, we find $2-3\sigma$ significance of rejection of featureless PPS, wiggles in WWI-C and WWI$'$ show 
just $3\sigma$ C.L. WWI-D on the other hand 
has highest probability of detection where the amplitude of wiggles reject {\it zero} amplitude at much more than 3$\sigma$. 
Since WWI-D has oscillations continuing to small 
scales, CORE is expected to capture its fine oscillations at high significance with better small scale measurements. 
In all the cases we find $<2\sigma$ significance for 
the suppression at large scale primordial power since cosmic variance is expected to limit 
our ability to detect such large scale signals \footnote{However, we must mention that, here we are only considering 
the local best fits to the Planck 2015 data obtained in the WWI framework.
It is entirely possible that even within this framework there 
are wide features with higher amplitude that are still allowed by the Planck data and which would be probed 
at higher significance with CORE. Our analysis does not rule out such possibilities.}.

We have observed a consistent reduction of the capability of future CMB polarization data to probe features 
at large scales with respect to previous literature, as also noticed in~\cite{ballardinietal}.
The degraded significance of the feature generated by a step in the potential (similar to WWI-A)
as compared to~\cite{mortonson} can be attributed to two facts. 
In this paper we use CORE noise sensitivity rather a cosmic variance ideal experiment as in~\cite{mortonson}. 
Secondly, wiggles in WWI-A are 22\% 
lower in amplitude compared to best fit potential-step-like feature obtained in WMAP. 
Although the Planck measurements have somewhat reduced the suggestion of large scale features, 
the combination of data from future galaxy surveys with CMB observations is an interesting avenue to probe the primordial origin 
of the deviations from the $\Lambda$CDM best-fit in Planck temperature data~\cite{ballardinietal}.

To probe extended reionization histories, we use a Piecewise Cubic Hermite Interpolating Polynomial 
to model the free electron fraction as a function of redshift. This is a more economic extension of the instantaneous reionization 
history than the PCA~\cite{PCA} approach which might be useful given the decrement in the value of the optical depth with 
Planck~\cite{Planck:2015Like,Aghanim:2016yuo}. Assuming
completion of hydrogen reionization by $z<5.5$, in agreement with current astrophysical observations, 
and a monotonic decrease of the electron fraction, we confront this model against Planck 2015 data with one extra parameter compared 
to the Tanh-Reion case. Although Poly-Reion does not encompass the Tanh-Reion for all optical depth values, 
we find marginal improvement in fit compared to the standard case. In 
this parametrization we could constrain the duration of reionization 
(redshifts between 10\% to 99\% reionization) within 1 to 15 redshifts with Planck data at 2$\sigma$ C.L. CORE spectra 
is expected to improve the constrain three times compared to Planck data.

Apart from proposing Poly-Reion as suitable extended reionization model, we also make use of it in exploring 
the possible confusion between primordial features and reionization histories. 
Although this monotonic reionization history is not able to reproduce 
the primordial features in polarization to 
a great extent, we show that large scale suppression 
from WWI-A can be slightly obscured by the presence of the reionization history. 
The significance of feature when WWI-A is used as fiducial spectra, decrease 
from $2-3\sigma$ to $1-2\sigma$ if Poly-reion is used as the model of reionization. 
Further investigations show that a more complex but realistic and monotonic reionization history
can not create much confusion with the inflationary features given CORE sensitivity where temperature,
polarization and lensing likelihoods are used for parameter constraints.

%%%%%%%%%%%%%%%%%%%%%%%%%%%%%%%%%%%%%%%%%%%%%%%%%%%%%%%%%%%%%%%%%%%%%%%%%%%%%%%
\section*{Acknowledgments}
The authors would like to acknowledge the use of APC cluster (\href{https://www.apc.univ-paris7.fr/FACeWiki/pmwiki.php?n=Apc-cluster.Apc-cluster}{https://www.apc.univ-paris7.fr/FACeWiki/pmwiki.php?n=Apc-cluster.Apc-cluster}).
DKH and GFS acknowledge Laboratoire APC-PCCP, Universit\'e Paris Diderot and Sorbonne Paris Cit\'e (DXCACHEXGS)
and also the financial support of the UnivEarthS Labex program at Sorbonne Paris Cit\'e (ANR-10-LABX-0023 and ANR-11-IDEX-0005-02).
DP, MB and FF acknowledge financial support by ASI Grant 2016-24-H.0 and 
partial financial support by the ASI/INAF Agreement I/072/09/0 for the Planck LFI Activity of Phase E2.
MB acknowledge the support from the South African SKA Project.
AS would like to acknowledge the support of the National Research Foundation of Korea (NRF-2016R1C1B2016478).
AAS was partially supported by the grant RFBR 17-02-01008 and by the Russian Government Program of Competitive 
Growth of Kazan Federal University.
% \clearpage
%%%%%%%%%%%%%%%%%%%%%%%%%%%%%%%%%%%%%%%%%%%%%%%%%%%%%%%%%%%%%%%%%%%%%%%%%%%%%%%

%%%%%%%%%%%%%%%%%%%%%%%%%%%%%%%%%%%%%%%%%%%%%%%%%%%%%%%%%%%%%%%%%%%%%%%%%%%%%%%
\end{document}